\pdfoutput=1

\documentclass[11pt,twoside,a4paper,cmspaper,final,collab]{cms-tdr}
\def\svnVersion{210296}\def\cmsCernNo{2013-086}\def\cmsCernDate{\today}\def\cmsMessage{Published in Physical Review D as \href{http://dx.doi.org/10.1103/PhysRevD.88.052001}{\doi{10.1103/PhysRevD.88.052001.}}}

\begin{document}\cmsNoteHeader{QCD-11-010}

\hyphenation{had-ron-i-za-tion}
\hyphenation{cal-or-i-me-ter}
\hyphenation{de-vices}

\RCS$Revision: 196021 $
\RCS$HeadURL: svn+ssh://svn.cern.ch/reps/tdr2/papers/QCD-11-010/trunk/QCD-11-010.tex $
\RCS$Id: QCD-11-010.tex 196021 2013-07-12 15:10:50Z alverson $
\providecommand{\Vz}{\ensuremath{\cmsSymbolFace{V}^0}\xspace}
\newlength\cmsFigWidth
\ifthenelse{\boolean{cms@external}}{\setlength\cmsFigWidth{0.95\columnwidth}}{\setlength\cmsFigWidth{0.7\textwidth}}
\ifthenelse{\boolean{cms@external}}{\providecommand{\cmsLeft}{top}}{\providecommand{\cmsLeft}{left}}
\ifthenelse{\boolean{cms@external}}{\providecommand{\cmsRight}{bottom}}{\providecommand{\cmsRight}{right}}
\cmsNoteHeader{QCD-11-010} 
\title{Measurement of neutral strange particle production in the underlying event in proton-proton collisions at \texorpdfstring{$\sqrt{s} = 7\TeV$}{sqrt(s)=7 TeV}}

\date{\today}

\abstract{

Measurements are presented of the production of primary $\PKzS$ and $\PgL$ particles in proton-proton
collisions at $\sqrt{s} = 7\TeV$ in the region transverse to the leading charged-particle jet in each
event. The average multiplicity and average scalar transverse momentum
sum of $\PKzS$ and $\PgL$
particles measured at pseudorapidities $\abs{\eta} < 2$ rise with increasing charged-particle jet \PT in
the range 1--10\GeVc and saturate in the region 10--50\GeVc. The rise and saturation of the
strange particle yields and transverse momentum sums in the underlying event are similar to those observed for
inclusive charged particles, which confirms the impact-parameter
picture of multiple parton interactions.
The results are compared to recent tunes of the \PYTHIA Monte Carlo event generator. The \PYTHIA
simulations underestimate the data by 15--30\% for
$\PKzS$ mesons and by about 50\% for $\PgL$
baryons, a deficit similar to that observed for the inclusive
strange particle production in non-single-diffractive proton-proton
collisions. The constant strange- to charged-particle activity ratios
with respect to the leading jet \PT and the similar trends for
mesons and baryons indicate that the multiparton-interaction dynamics is decoupled from parton
hadronization, which occurs at a later stage.
}

\hypersetup{%
pdfauthor={CMS Collaboration},%
pdftitle={Measurement of neutral strange particle production in the underlying
event in proton-proton collisions at sqrt(s) = 7 TeV},%
pdfsubject={CMS},%
pdfkeywords={CMS, QCD, physics}}

\maketitle 

\section{Introduction}
\label{sec:intro}

This paper describes a measurement of the
production of primary $\PKzS$ mesons,
$\PgL$ and $\PagL$ baryons in the underlying event in proton-proton ($\Pp\Pp$) collisions at a centre-of-mass
energy of 7\TeV with the Compact Muon Solenoid (CMS) detector at the Large
Hadron Collider (LHC).

In the presence of a hard process, characterized by particles
or clusters of particles with large transverse
momentum $\PT$ with respect to the beam direction, the final state of
hadron-hadron interactions can be described
as the superposition of several contributions: the partonic hard
scattering, initial- and final-state radiation,
additional ``multiple partonic interactions'' (MPI) and ``beam-beam
remnants'' (BBR) interactions.
The products of initial- and final-state radiation, MPI and BBR, form the ``underlying event'' (UE).

In this paper, the UE properties are analyzed with reference to the
direction of the highest-$\PT$ jet reconstructed from charged primary particles
(leading charged-particle jet). This leading jet is expected to reflect the direction of the
parton produced with the highest transverse momentum in the hard interaction.
Three distinct topological regions in the hadronic final state are defined
in terms of the azimuthal angle $\Delta\phi$ between the directions
of the leading jet and that of any particle in the event.
Particle production in the ``toward" region, $\abs{\Delta\phi} < 60^\circ$, and in
the ``away" region, $\abs{\Delta\phi}> 120^\circ$, is expected to be
dominated by the hard parton-parton scattering.
The UE structure can be best studied in the ``transverse"
region, $60^\circ < \abs{\Delta\phi}  < 120^\circ$~\cite{CMS-PAS-QCD-10-010,Field:2010su}.

Studies of the UE activity in charged primary particles in
proton-proton collisions at different centre-of-mass energies
have been published by the ATLAS~\cite{Aad:2010fh} and
CMS~\cite{Khachatryan:2010pv, CMS-PAS-QCD-10-010, CMS-PAS-QCD-11-012} collaborations.
Observables such as the average multiplicity of charged primary
particles per event, hereafter referred to as ``average rate'',
and the average scalar sum of primary particle $\PT$ per event, hereafter referred to as
``average $\PT$ sum'', have been measured in the transverse
region. These quantities exhibit a steep rise with increasing charged-particle jet $\PT$ up to a
value that depends on the proton-proton centre-of-mass energy
(around $10\GeVc$ for $\Pp\Pp$ collisions at 7\TeV), followed by
a slow rise.
Within the MPI framework, a hard jet is likely to be produced in
collisions with a small impact parameter between the colliding
protons, consequently resulting in large MPI
activity~\cite{Sjostrand:1986ep,Frankfurt:2010ea}.
The MPI activity saturates at values of the hard scale typical of central
collisions.

The present analysis considers identified neutral strange particles
($\PKzS$, $\PgL$, and $\PagL$) as additional probes to study
the underlying event. Unless stated otherwise, $\PgL$ and
$\PagL$ baryon data are merged and referred to as
$\PgL$ baryon data.
The production of primary
$\PKzS$ and $\PgL$ particles in the transverse region at $\sqrt{s} = 7\TeV$
is studied as a function of the scale of the hard process.
Fully corrected average rates and $\PT$ sums of primary
$\PKzS$ mesons and $\PgL$ baryons,
as well as ratios to the charged
primary-particle rates and $\PT$ sums
are compared to simulations. This
analysis complements the studies of strangeness production
in minimum-bias events at $\sqrt{s} = 7\TeV$ published by the
ALICE~\cite{Abelev:2012jp, Chinellato:2012jj},
ATLAS~\cite{Aad:2011hd} and CMS~\cite{Khachatryan:2011tm} collaborations.
Comparisons of non-single diffractive data~\cite{Khachatryan:2011tm}
with predictions made
with the \PYTHIA6~\cite{Sjostrand:2006za} and
\PYTHIA8~\cite{Sjostrand:2007gs}
Monte Carlo event generators have
shown that the latter largely underestimate the data,
\eg by 30\% for $\PKzS$
production and 50\% for $\PgL$ production
at $\sqrt{s} = 7\TeV$ for \PYTHIA6 tune {D6T}~\cite{Field:2010su, Field:2008zz},
with little improvement for more recent tunes.

The simulations are performed with versions of \PYTHIA
that include MPI. The most recent versions have been tuned to reproduce the UE
activity observed with primary charged particles at the LHC at 0.9\TeV
and 7\TeV centre-of-mass energies. The parameters describing strangeness
production, however, have not been tuned to LHC data yet.
All Monte Carlo samples used in this paper
have been generated with the default values of these parameters.

Recent literature~\cite{Buckley:2009bj,Drescher:2001hp,Abreu:2007kv} discussing the
tuning of the strangeness suppression parameters in commonly-available
generators is limited.
A tuning of the \PYTHIA6 parameters to LEP, SLD and Tevatron
data performed with the \textsc{professor} program~\cite{Buckley:2009bj}
produced best-fit parameters in disagreement with the current \PYTHIA
default parameters. The resulting predicted strange meson and baryon
production rates given in the Appendix of Ref.~\cite{Buckley:2009bj},
however, do not agree well with the data used for the tuning.
Other attempts to describe strange particle production in $\Pp\Pp$ collisions
are discussed in Refs.~\cite{Drescher:2001hp,Abreu:2007kv}. The present paper focuses on the comparison
with \PYTHIA.

The outline of this paper is the following. In Section~\ref{sec:data},
the experimental conditions are described, along with the data sets,
the simulation and the analysis technique. In Section~\ref{sec:systematics},
the systematic uncertainties are summarized. The results are discussed
in Section~\ref{sec:results}, and conclusions are drawn in
Section~\ref{sec:conclusions}.

\section{Experimental setup, data sets and data analysis}
\label{sec:data}

The central feature of CMS is a
superconducting solenoid of 6\unit{m} internal diameter.
Within the superconducting solenoid volume are a silicon pixel and
strip tracker, a lead tungstate crystal electromagnetic calorimeter,
and a brass/scintillator hadron calorimeter.
Muons are measured in gas-ionization detectors embedded in the flux-return
yoke. Extensive forward calorimetry complements the coverage provided
by the barrel and endcap detectors.
CMS uses a right-handed coordinate system, with the origin at the
nominal interaction point, the $x$ axis pointing to the centre of the
LHC, the $y$ axis pointing up (perpendicular to the LHC plane), and
the $z$ axis along the anticlockwise-beam direction. The polar angle
$\theta$ is measured from the positive $z$ axis and the azimuthal
angle $\phi$ is measured in the $x$-$y$ plane.
The tracker measures charged particles within the pseudorapidity
range $\abs{\eta}< 2.5$, where $\eta = -\ln (\tan(\theta / 2)  )$.
It consists of 1440 silicon pixel and 15\,148
silicon strip detector modules and is located in the $3.8\unit{T}$ field
of the superconducting solenoid. For the charged particles of interest in this
analysis, the transverse momentum resolution is relatively constant
with $\PT$, varying from 0.7\% at $\eta =0$ to 2\% at $\abs{\eta}=2$.
The transverse and longitudinal impact parameter resolutions,
$\sigma_{d_0}$ and $\sigma_{d_z}$, respectively, depend
on $\PT$ and on $\eta$, ranging from $\sigma_{d_0} =
400\micron$ and $\sigma_{d_z} = 1000\micron$ at
$\PT = 0.3\GeVc$ and $\abs{\eta}>1.4$ to $\sigma_{d_0} = 10\micron$
and $\sigma_{d_z} = 30\micron$ at $\PT = 100\GeVc$ and $\abs{\eta}<0.9$.
A more detailed description of the CMS detector can be found in Ref.~\cite{CMS:2008zzk}.

\subsection{Event selection, data sets and Monte Carlo simulation}
\label{ssec:eventsel}

The event selection is identical to the one described
in~\cite{CMS-PAS-QCD-10-010}, unless explicitly stated otherwise.
Minimum-bias events were triggered by
requiring coincident signals in beam scintillator
counters located on both sides of the experiment and covering the pseudorapidity range $3.23 < \abs{\eta} <
4.65$, and in the beam pick-up devices~\cite{CMS:2008zzk}.
Events were then recorded with a prescaled trigger requiring the presence
of at least one track segment in the pixel detector with $\PT > 200\MeVc$. The trigger conditions are applied to both data and simulated
samples. The trigger efficiency for the events selected in the
analysis is close to 100\% and no bias from the trigger selection is
found.

The data used in this analysis were collected in early 2010 when
pile-up (multiple $\Pp\Pp$ collisions per proton
bunch crossing) was very low.
Selected events are required to contain a single reconstructed
primary vertex, a condition that rejects about 1\% of the events
satisfying all the other selection criteria.
The primary vertex is fit with an adaptive
algorithm~\cite{Fruhwirth:2007hz} and must have at least four tracks,
a transverse distance to the beam line smaller than 2\unit{cm},
and a $z$ coordinate within 10\unit{cm} of the nominal interaction point.

Events are required to contain a track-jet with
reconstructed $\PT > 1\GeVc$ and $\abs{\eta} < 2$. Track-jets are
reconstructed from the tracks of charged particles, with the anti-\kt
algorithm~\cite{Cacciari:2005hq,Cacciari:2008gp} and a clustering radius $\Delta R = 0.5$,
where $\Delta R = \sqrt{(\Delta \eta)^2 + (\Delta \phi)^2}$.
The tracks are required to be well reconstructed, to have $\PT > 500\MeVc$, $\abs{\eta} < 2.5$, and
to be consistent with originating from the primary vertex.
More details on the track selection can be found
in~\cite{CMS-PAS-QCD-10-010}.
The reconstructed track-jet $\PT$ is the magnitude of the vector sum
of the transverse momenta of the tracks in the jet.
The leading track-jet $\PT$ is corrected for detector response
(track finding efficiency and $\PT$ measurement) with detailed
simulations based on \GEANTfour\cite{Agostinelli:2002hh}, which have
been extensively validated with data~\cite{CMS-PAS-TRK-10-005,CMS-PAS-TRK-10-002,Khachatryan:2010pw}.
This correction is approximately independent of the track-jet $\PT$ and
$\eta$ and its average value is 1.01. The leading corrected track-jet
is referred to as the leading charged-particle jet.

The \PYTHIA versions considered all include MPI.
The tunes used are the \PYTHIA6 D6T tune~\cite{Field:2010su, Field:2008zz}
and the \PYTHIA8 tune 1~\cite{Sjostrand:2007gs}, which have not been tuned
to the LHC data, and
the \PYTHIA6 {Z1}~\cite{Field:2010bc} and {Z2*} tunes.
The two latter \PYTHIA6 tunes, as well as \PYTHIA8, include $\PT$ ordering
of the parton showers, and a new model~\cite{Corke:2010yf} where
MPI are interleaved with parton showering.
\PYTHIA8 includes hard diffraction in addition to the new MPI
model. The parton distribution functions
used for \PYTHIA6 D6T and \PYTHIA8 tune 1 are the CTEQ6L1 and
CTEQ5L sets, respectively. The {Z1} tune uses the CTEQ5L
parton distribution set, whereas {Z2*} is updated to
CTEQ6L1~\cite{Pumplin:2002vw} and retuned to the underlying event
activity at $7\TeV$ from Ref.~\cite{CMS-PAS-QCD-10-010}
with the \textsc{professor} tool~\cite{Buckley:2009bj}. The simulated data are generated with \PYTHIA6 version 6.422
for tunes \textsc{D6T} and \textsc{Z1}, version 6.424 for tune {Z2*}, and version
8.135 for \PYTHIA8 tune 1.

Simulated primary stable charged particles with a proper lifetime $c\tau > 1\unit{cm}$ are clustered into jets with the anti-\kt
algorithm ($\Delta R = 0.5$). The average rates and scalar $\PT$ sums of
simulated primary $\PKzS$ and $\PgL$ particles
are computed within the transverse region of the leading simulated charged-particle jet.

A data sample of 11 million events with at least one
charged-particle jet with $\PT > 1\GeVc$ and $\abs{\eta} < 2$ is
analysed. The corresponding
numbers of simulated events are 22 million for \PYTHIA6 D6T
and 5 million for \PYTHIA6 Z1, Z2* and \PYTHIA8 tune 1.
Corrections for detector effects and background are
estimated with the \PYTHIA6 D6T sample, while the modeling of
the underlying event is studied with all the tunes mentioned.

The reconstruction of the leading charged-particle jet results in a bias
in the measured average rates and $\PT$ sums in the transverse
region. The value of this bias ranges from $+5$\% to $+10$\% for charged-particle jet $\PT$ below
$10\GeVc$, and is consistent with zero for larger $\PT$ values.
It is caused by events in which the leading jet formed by
primary charged particles is not reconstructed as the leading charged-particle jet
because of tracking inefficiencies, and a sub-leading jet is thus
reconstructed as the leading jet. This results in a reconstructed transverse
region shifted in $\phi$.
The correction for this bias is obtained from the detailed Monte Carlo
simulations of the detector response described above.

The primary vertex selection causes a small overestimate of the UE
strangeness activity at low charged-particle jet $\PT$, at most 5\%
for charged-particle jet $\PT = 1\GeVc$. This is because the requirement that at least four tracks
be associated to the primary vertex enriches the sample
in events with higher UE activity when the charged-particle jets have very low
multiplicity. This bias is corrected by means of detailed simulations as described in
Section~\ref{sec:systematics}.

\subsection{Selection of primary \texorpdfstring{$\Vz$}{V0} candidates and analysis strategy}
\label{ssec:strategy}

The neutral strange particles $\PKzS$, $\PgL$ and
$\PagL$, hereafter generically called $\Vz$s,
are identified by means of their characteristic decay
topology: a flight distance of several centimeters before decay, two tracks of
opposite charge emerging from a secondary vertex, and an invariant
mass consistent with that of a $\PKzS$ meson or a $\PgL$
baryon. The $\Vz$ momentum vector is further required to be collinear with
the vector joining the primary and secondary vertices, in order to select primary
particles.

The $\Vz$ candidates are reconstructed by the standard CMS offline event reconstruction program~\cite{Khachatryan:2010pw}.  Pairs of oppositely-charged tracks with at least 3 hits in the CMS tracker and with a non-zero transverse impact parameter with respect to the beam line are selected (the transverse impact parameter divided by its uncertainty is requested to be larger than 1.5). Pairs of tracks with a distance of closest approach to each other smaller than 1\unit{cm} are fit to a common secondary vertex, and those with a vertex fit $\chi^2$ smaller than 7 and a significant distance between the beam line and the secondary vertex (transverse flight distance divided by its uncertainty larger than 8) are retained.

Well-reconstructed $\Vz$ candidates are selected by applying
cuts on the pseudorapidity and transverse momentum of the decay tracks ($\abs{\eta} < 2.5$, $\PT > 300\MeVc$), of the $\Vz$ candidate ($\abs{\eta} < 2$; $\PT >600\MeVc$ for $\PKzS$ mesons, $\PT >1.5\GeVc$ for $\PgL$ baryons) and on the $\Vz$ transverse flight distance ($>$1\unit{cm} from the beam line). A kinematic fit is then performed on the candidates to further purify the sample of primary strange particles. The fit includes a secondary vertex constraint,
a mass constraint, as well as the constraint that the $\Vz$ momentum points away from the primary vertex. All three hypotheses ($\PKzS \to \pi^+\pi^-$, $\PgL \to \Pp \pi^-$ and $\PagL \to \Pap \pi^+$) are tested for each candidate and the most probable hypothesis is considered. Candidates with a kinematic-fit probability larger than 5\% are retained.
\begin{figure*}[htbp]
\begin{center}
\includegraphics[width=0.45\textwidth]{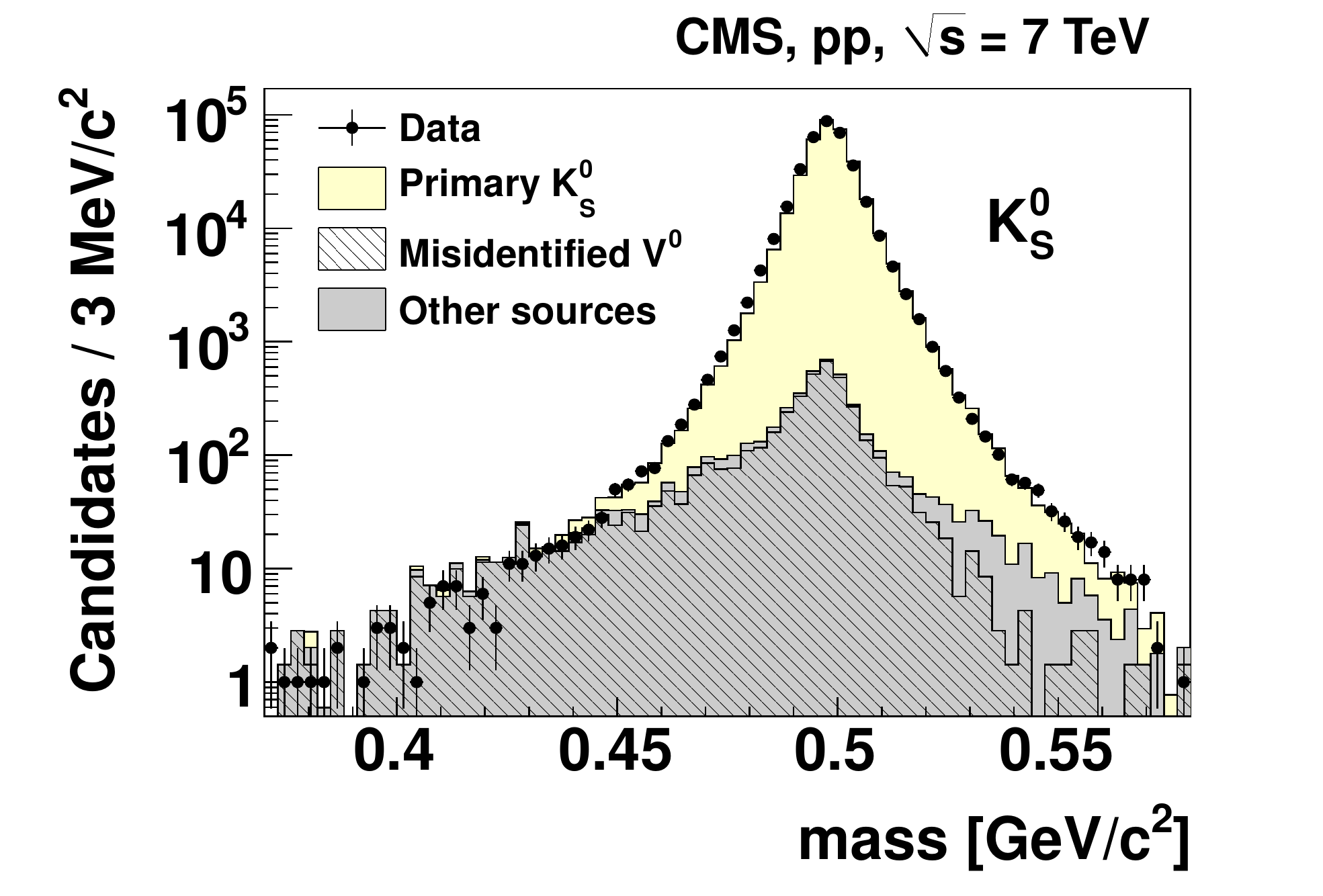}
\includegraphics[width=0.45\textwidth]{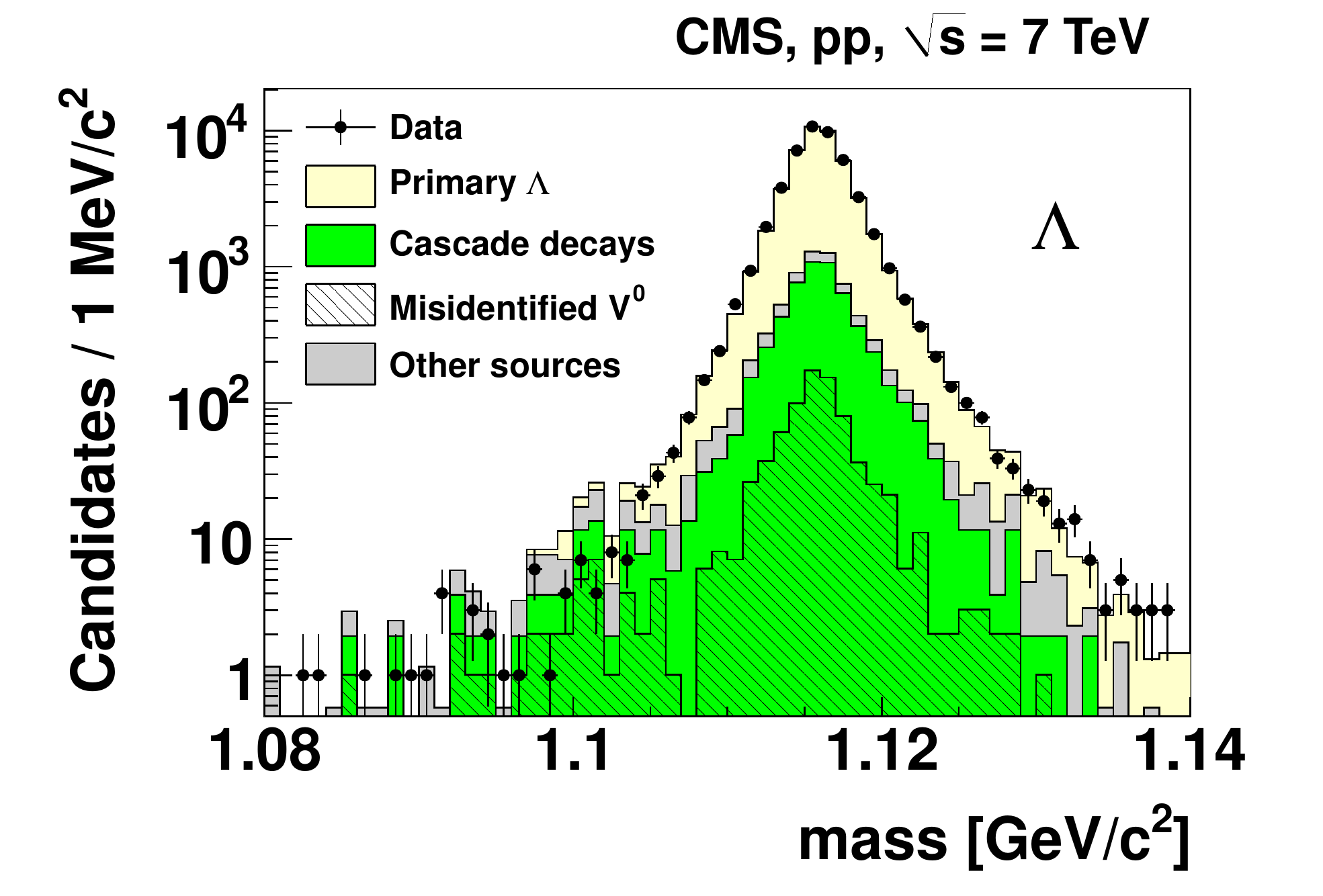}
\includegraphics[width=0.45\textwidth]{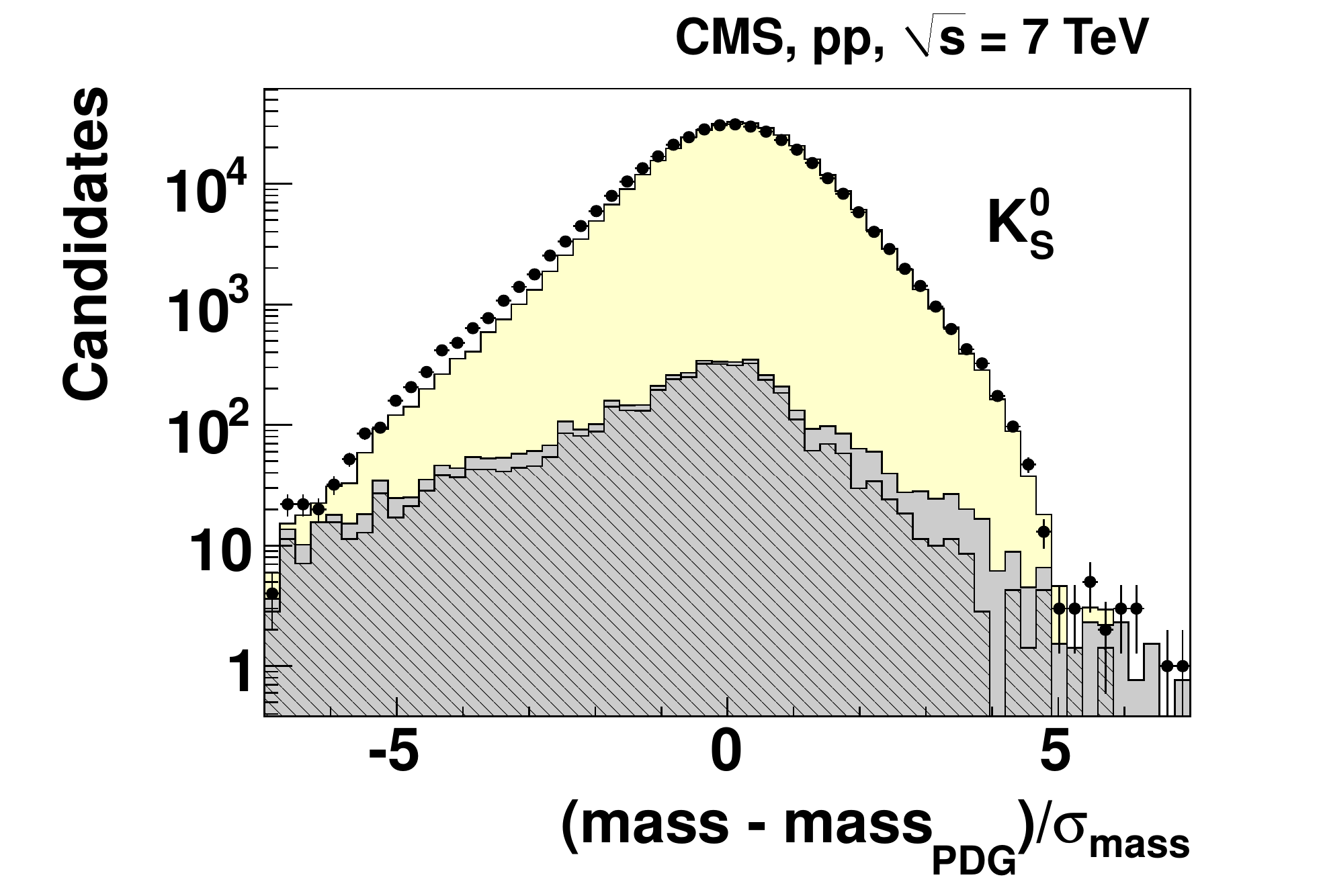}
\includegraphics[width=0.45\textwidth]{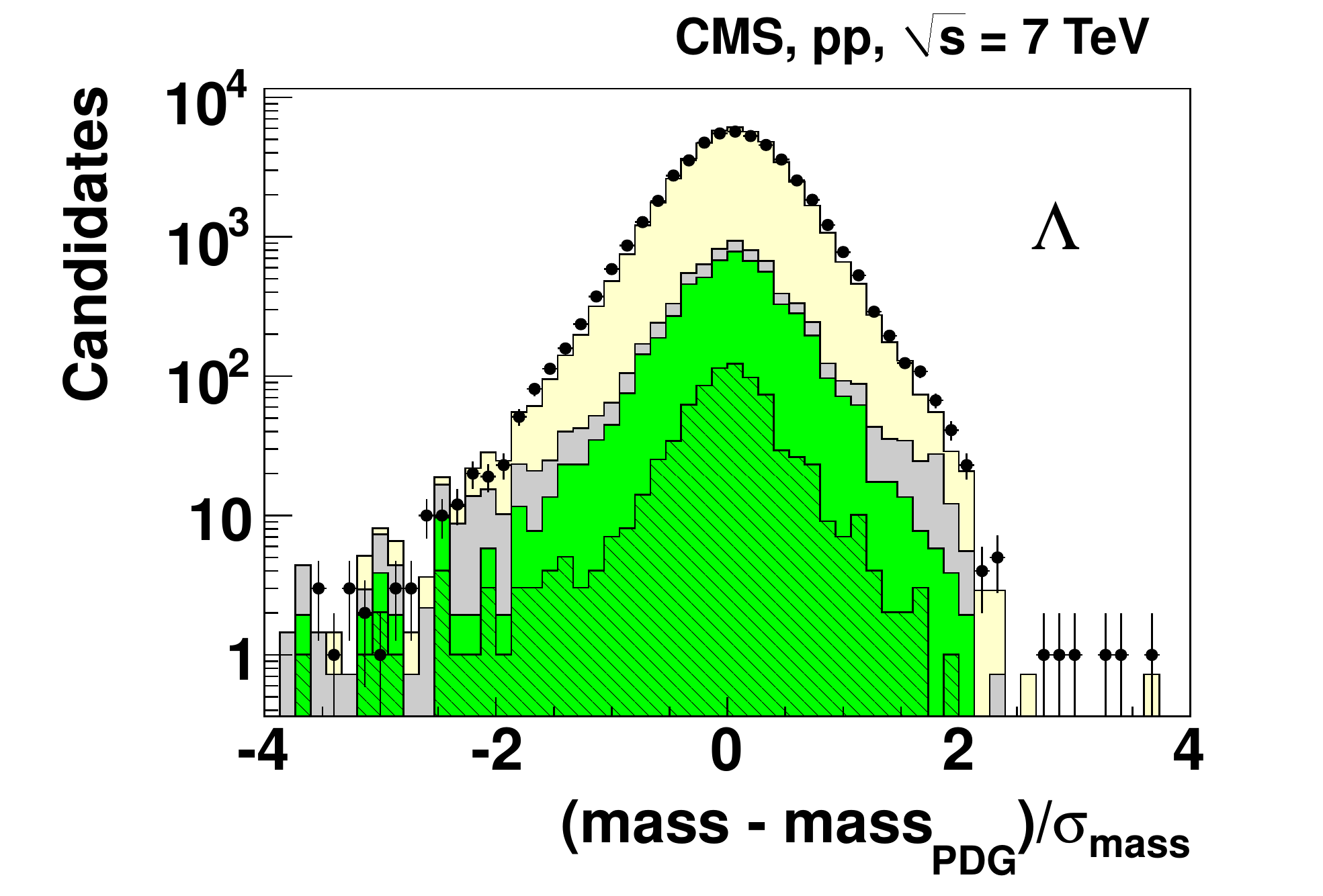}
\caption{Distributions of invariant mass and invariant-mass pull for the most probable particle type hypothesis determined by the kinematic fit. The accepted $\PKzS$ and $\PgL$ mass values from Ref.\cite{pdg2012} are denoted as $\mathrm{mass_{PDG}}$. The black points indicate the data. The histograms show the backgrounds (hatched: misidentified $\Vz$; green: non-primary $\PgL$ from $\Xi$ and $\Omega$ cascade decays; grey: other sources) and the signal (yellow) as predicted by \PYTHIA6 D6T\@.
The \PYTHIA prediction is normalized to the data.}
\label{fig:v0mass}
\end{center}
\end{figure*}

Since simulations enter in the determination of the $\Vz$ selection efficiency and purity, a good description of the distributions of the kinematic-fit input variables is important. The distributions of the invariant mass of the $\Vz$ candidates for the most probable particle type hypothesis are shown in Fig.~\ref{fig:v0mass}, together with the distributions of the invariant-mass pull. The invariant-mass pull is the difference between the reconstructed mass and the accepted $\Vz$ mass value~\cite{pdg2012}, divided by the uncertainty on the reconstructed mass calculated from the decay track parameter uncertainties. The signal and background fractions are shown as predicted by \PYTHIA6 D6T\@.
The backgrounds in the  $\PKzS$ sample are mostly misidentified $\PgL$
baryons. Backgrounds in the $\PgL$ sample are mostly non-primary
$\PgL$ baryons from cascade decays of $\Xi$ and $\Omega$ baryons, plus
contributions from misidentified $\PKzS$ mesons and converted
photons. In general, the simulation agrees with the data. As an
example, the average mass values for $\PKzS$ mesons ($\PgL$ baryons) are 0.4981\GeVcc (1.116\GeVcc)
in the simulation and 0.4977\GeVcc (1.116\GeVcc) in
the data; the corresponding R.M.S. values for the mass pull distributions are 1.17 (0.512)
in the simulation and 1.23 (0.531) in the data.
For $\PKzS$
candidates, the data show larger tails than the simulation at mass
pull values below ($-$2). The presence of a similar tail in the
component shown as the hatched histogram of the simulated distribution indicates that this excess is
due to a larger contribution from misidentified baryons in the data
compared to the simulation. This is accounted for in the background estimation as described below.

\begin{figure*}[htbp]
\begin{center}
\includegraphics[width=0.45\textwidth]{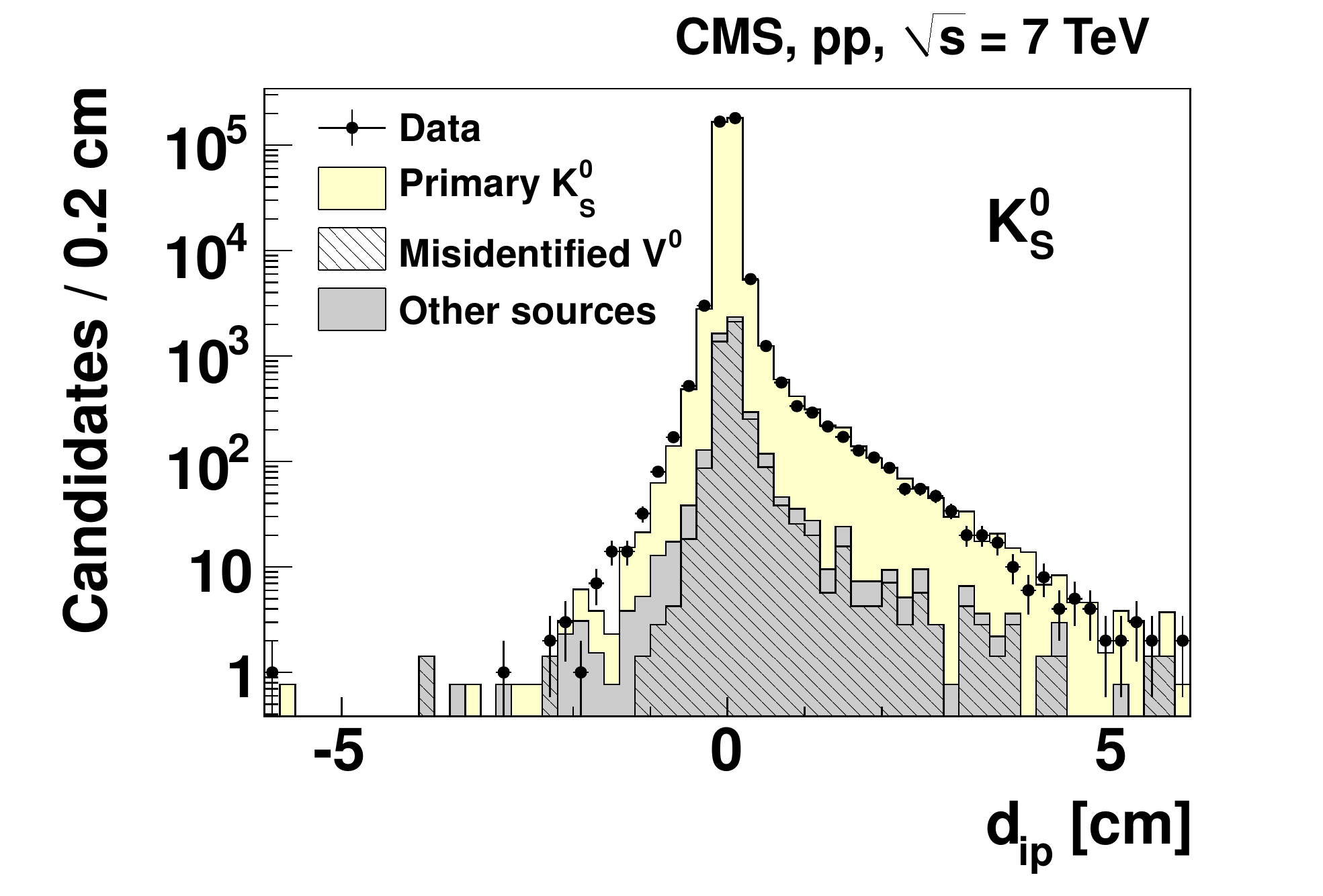}
\includegraphics[width=0.45\textwidth]{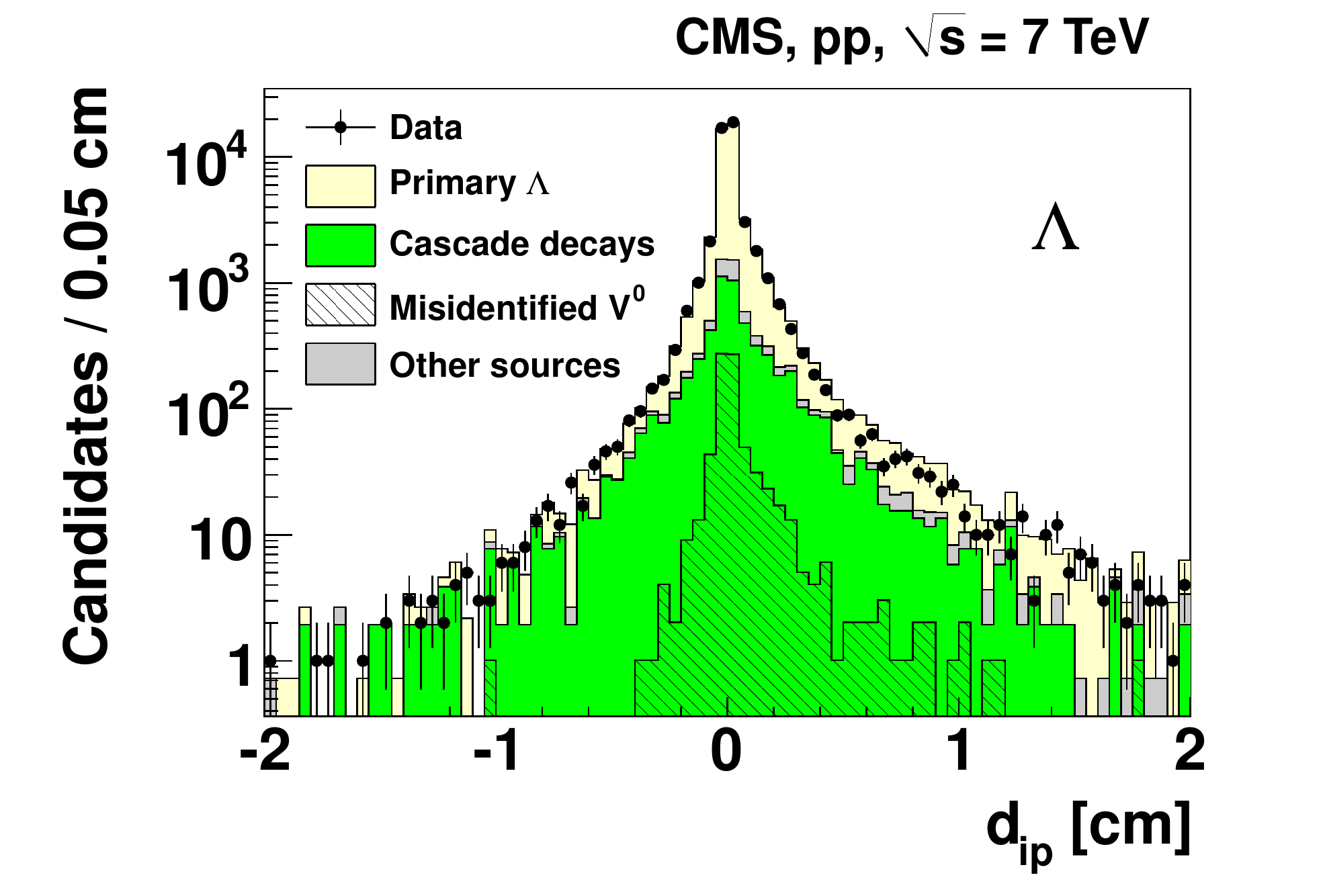}
\includegraphics[width=0.45\textwidth]{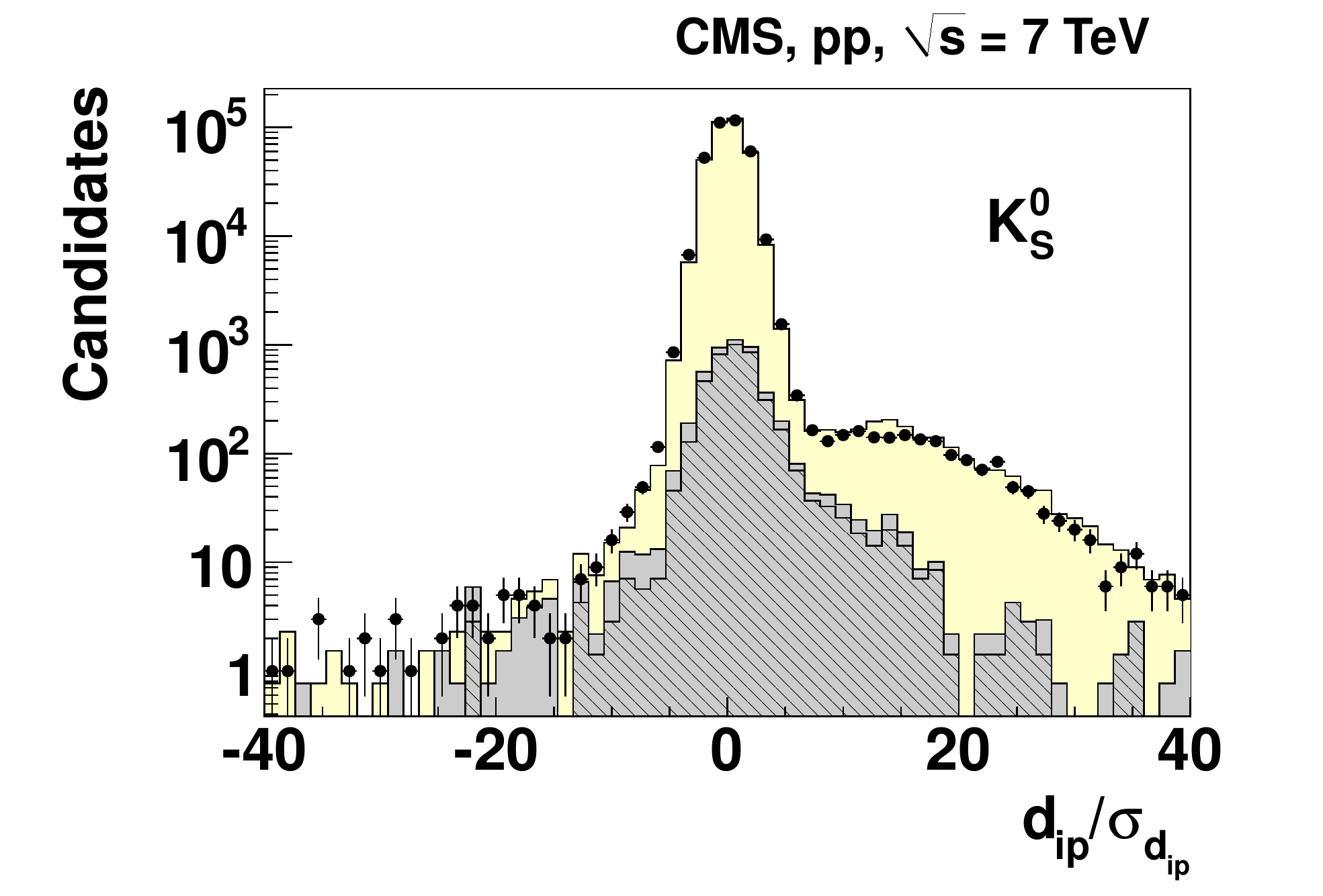}
\includegraphics[width=0.45\textwidth]{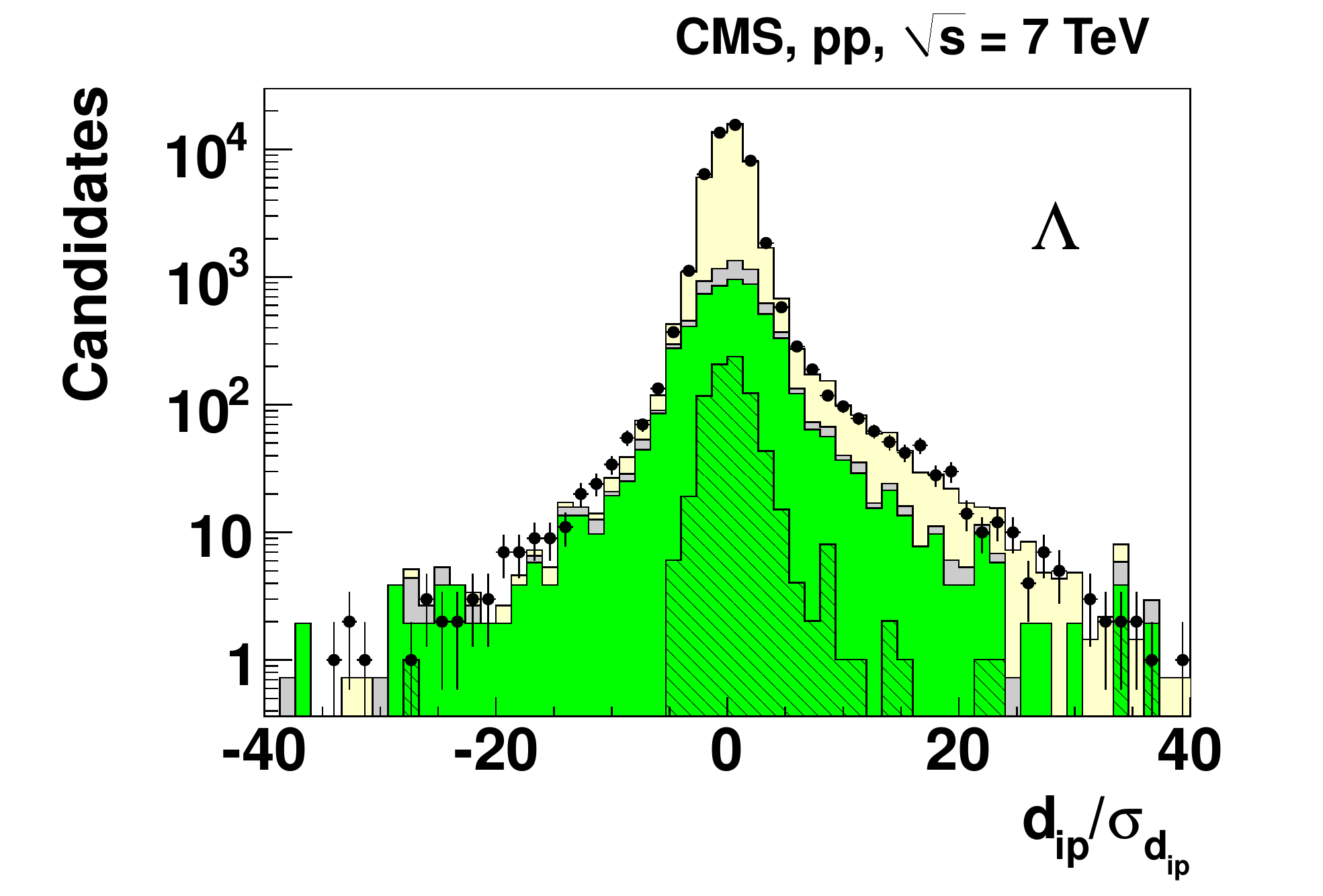}
\caption{Distributions of the signed impact parameter $d_\mathrm{ip}$ with respect to the primary vertex, and the corresponding pull distributions, for the most probable particle type hypothesis determined by the kinematic fit. The black points indicate the data. The histograms show the backgrounds (hatched: misidentified $\Vz$; green: non-primary $\PgL$ from $\Xi$ and $\Omega$ cascade decays; grey: other sources) and the signal (yellow) as predicted by \PYTHIA6 D6T\@.
The \PYTHIA prediction is normalized to the data.}
\label{fig:v0dist}
\end{center}
\end{figure*}

The pointing requirement constrains the signed impact parameter $d_\mathrm{ip}$ of the $\Vz$ with respect to the primary vertex. This variable is defined as the distance of closest approach of the $\Vz$ trajectory to the primary vertex, and its sign is that of the scalar product of the $\Vz$ momentum and the vector pointing from the primary vertex to the point of closest approach. The distributions of the signed impact parameter are shown in Fig.~\ref{fig:v0dist} together with the distributions of the corresponding pull, defined as $d_\mathrm{ip}$ divided by its uncertainty $\sigma_{d_\mathrm{ip}}$ calculated from the decay track parameter uncertainties. The quality of the description of the data by the simulation is good,
including the tails at positive impact parameter values. The large pulls for secondary $\PgL$ baryons from cascade decays allow the suppression of this background by means of the kinematic fit.

\begin{figure}[htbp]
\begin{center}
\includegraphics[width=\cmsFigWidth]{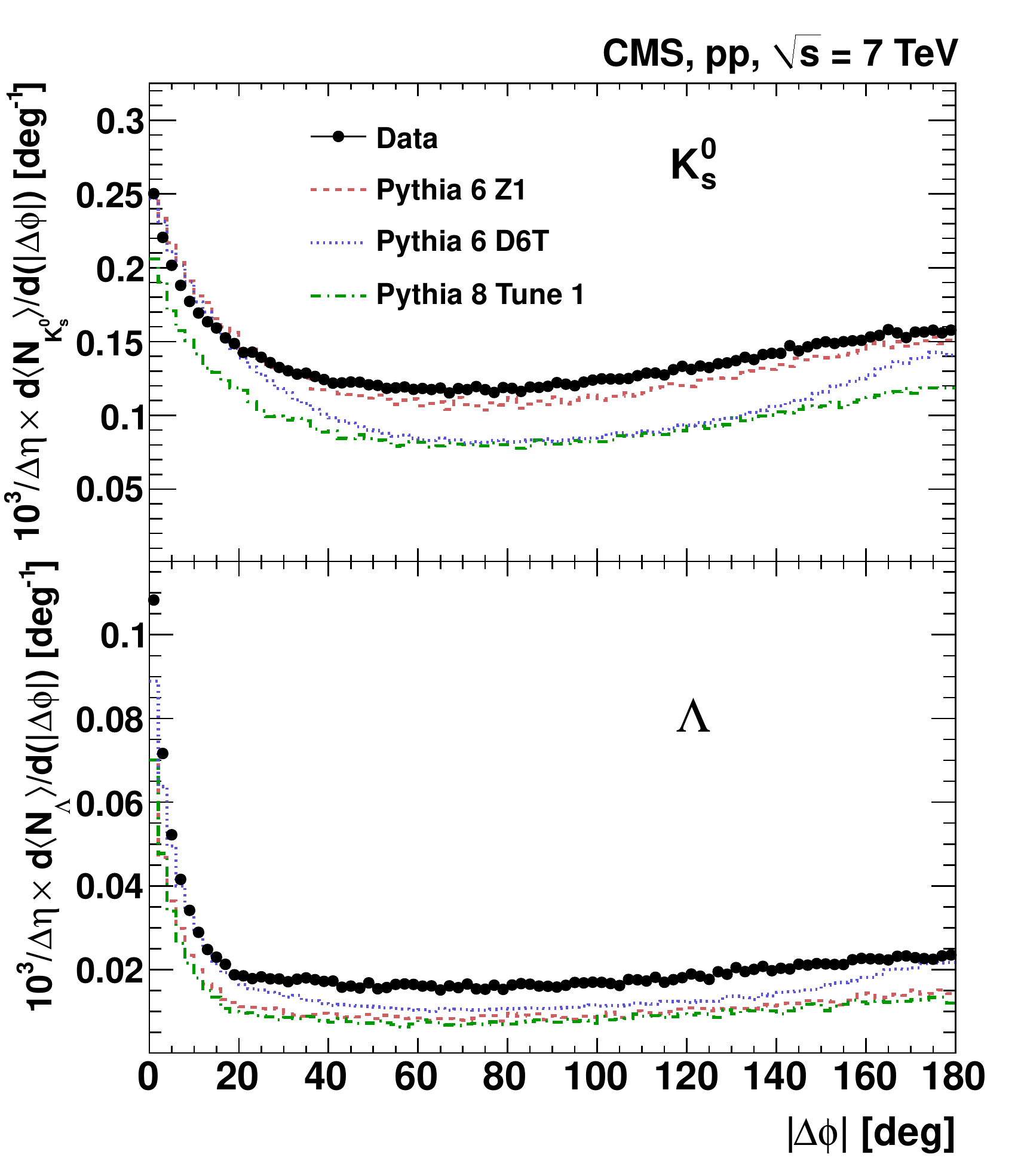}
\caption{Uncorrected average rate of selected $\Vz$ candidates per event, per degree and per unit pseudorapidity within $\abs{\eta} < 2$, as a function of the difference in azimuthal angle $\abs{\Delta\phi}$ between the $\Vz$ candidate and the leading charged-particle jet. Data and detailed simulation of minimum-bias events with different \PYTHIA tunes are shown for reconstructed charged-particle jet $\PT > 1\GeVc$. Top: $\PKzS$ candidates with $\PT > 600\MeVc$; bottom: $\PgL$ candidates with $\PT > 1.5\GeVc$.}
\label{fig:v0id_rates_dphi}
\end{center}
\end{figure}

The uncorrected average rates of reconstructed $\Vz$ candidates
passing the selection cuts per unit pseudorapidity are shown in
Fig.~\ref{fig:v0id_rates_dphi} as a function of the difference in azimuthal angle $\abs{\Delta\phi}$ between the $\Vz$ candidate and the leading charged-particle jet. Uncorrected data are compared to \PYTHIA events passed through the detailed detector simulation.
The dependence of the rates on $\abs{\Delta\phi}$ is qualitatively described by the \PYTHIA tunes considered. The simulation underestimates significantly the $\Vz$ rates in the transverse region. The peak at $\abs{\Delta\phi} \approx 0^\circ$ is more pronounced for baryons than for $\PKzS$ mesons. The simulation indicates that the harder $\PT$ cut applied to the baryon candidates is responsible for this feature; the distributions are similar when the same \PT cut is applied to both $\Vz$ types.

The backgrounds to the $\PKzS$ and $\PgL$ samples are estimated with
two methods. The first is based on simulation. Candidates not matched
to a generated primary $\Vz$ of the corresponding type are counted as
background. The \PYTHIA6 D6T sample is used. To account for the known
deficit of strange particles in the simulation (see Section~\ref{sec:intro}), the contribution from $\PKzS$ mesons misidentified as $\PgL$ baryons
is weighted by the ratio of $\PKzS$ rates measured in non-single diffractive events to those in \PYTHIA6 D6T, 1.39~\cite{Khachatryan:2011tm}. Similarly, the contribution from misidentified $\PgL$ baryons is weighted by a factor of 1.85, and the contribution arising from non-primary baryons from $\Xi$ and $\Omega$ decays is weighted by the ratio of the measured and simulated $\Xi$ production rates, 2.67~\cite{Khachatryan:2011tm}.

The second method is based on data. The signal and background contributions are extracted from a fit to the distribution of the kinematic fit $\chi^2$-probability, with signal and background shapes obtained from simulation. Apart from the background normalization, the measured and simulated pull distributions of the constrained variables (Figs.~\ref{fig:v0mass} and ~\ref{fig:v0dist}), as well as the measured and simulated $\chi^2$-probability distributions (not shown) are in good agreement.
These facts, as well as goodness-of-fit tests, validate the approach.

In both methods, the background is estimated as a function of the charged-particle jet $\PT$ for the rate measurements, and as a function of the $\Vz$ $\PT$ for the $\Vz$ $\PT$ spectra and the $\PT$ sum measurements. The background estimations from the two methods are in reasonable agreement, and exhibit the same dependence on the charged-particle jet and $\Vz$ $\PT$. The final background estimates are computed as the average of the results of the two methods, and the corresponding systematic uncertainties are taken as half the difference of the two results. The background fraction for $\PKzS$ increases from $(1.5 \pm 1.1)$\% at charged-particle jet $\PT = 1\GeVc$ to $(3.3 \pm 1.7$)\% at charged-particle jet $\PT = 10\GeVc$ and remains constant at higher charged-particle jet $\PT$. The background is $(8 \pm 2)$\% for baryons, independent of the charged-particle jet $\PT$.

The $\PKzS$ and $\PgL$ raw yields are corrected for purity (defined as 1 -- background fraction) as well as for acceptance and reconstruction efficiency. Each $\Vz$ candidate is weighted by the product of the purity times $\frac{1}{ A \times \epsilon}$, where $A$ denotes the acceptance of the cuts on the $\Vz$ transverse flight distance and on the $\PT$, $\eta$ of the decay particles, and $\epsilon$ denotes the reconstruction and selection efficiency for accepted $\Vz$ candidates. The product of acceptance times efficiency is computed in $\Vz$ $(\PT, \eta)$ bins from a sample of 50 million \PYTHIA6 D6T minimum-bias events passed through the detailed detector simulation. The average values of the product of acceptance and efficiency in this sample for $\PKzS$ mesons, $\PgL$ and $\PagL$ baryons within the kinematic cuts ($\abs{\eta} < 2$; $\PT >600\MeVc$ for $\PKzS$, $\PT >1.5\GeVc$ for $\PgL$ and $\PagL$) are 11.3\%, 8.4\% and 6.6\%, respectively, including the branching fractions $\mathcal{B}(\PKzS \to \pi^+\pi^-) = 69.2\%$ and $\mathcal{B}(\PgL \to \Pp \pi^-) = \mathcal{B}(\PagL \to \Pap \pi^+) = 63.9\%$~\cite{pdg2012}. The acceptance depends strongly on the $\Vz$ $\PT$, while the efficiency varies by a factor of about two in the $\Vz$ $\PT$ and $\eta$ ranges selected. The smaller efficiency for $\PagL$ baryons than for $\PgL$ baryons reflects the higher interaction cross section of antiprotons with the detector material compared to that of protons. The corrected $\PgL$ and $\PagL$ yields are found to be compatible when accounting for the systematic uncertainty due to the modelling of the antiproton cross section in the \GEANTfour version used~\cite{Aamodt:2010dx} (see Section~\ref{sec:systematics}).

The consistency of the correction method was checked by applying it to all other Monte Carlo samples and comparing the results to the known generated values. Further support to the correction procedure is provided by the fact that the
simulation reproduces well several key aspects of the data, most notably the reconstruction efficiency~\cite{CMS-PAS-TRK-10-005,CMS-PAS-TRK-10-002}
and the angular distributions of the $\Vz$ decay tracks as a function of the $\Vz$ $\PT$.
The reliability of the simulation for $\PKzS$ and $\PgL$ reconstruction was checked by comparing the lifetimes obtained from fits to the corrected proper time distributions with the world averages~\cite{Khachatryan:2011tm}. The stability of the results when varying the $\Vz$ selection cuts was also checked. The resulting overall contribution of the $\Vz$ reconstruction to the systematic uncertainty is given in Section~\ref{sec:systematics}.

\section{Systematic uncertainties}
\label{sec:systematics}

The main sources of systematic uncertainties are described below, with numerical values summarized in Table~\ref{tab:syst}.

\begin{description}

 \item[Leading charged-particle jet selection] The bias in rates and
   $\PT$ sums due to mismatches between the reconstructed and
   the simulated leading charged-particle jets is corrected by means of
   detailed simulations.
   The systematic uncertainty is estimated from the residual difference in rates and
   $\PT$ sums when the reconstructed and the simulated leading charged-particle jets
   are matched within $\Delta R = 0.3$.

  \item[Primary vertex selection]
    The bias caused by the requirement of a minimum track multiplicity
    at the primary vertex is corrected by means of detailed simulations of
    minimum-bias events with the \PYTHIA6 Z1 tune.
    The primary charged particle multiplicity in $7
   \TeV$ $\Pp\Pp$ collisions is well described by this tune~\cite{CMS-PAS-QCD-10-010}.
    The corresponding uncertainty is estimated from the spread of the
    corrections computed with \PYTHIA6 tunes \textsc{D6T},
    \textsc{Z1} and \PYTHIA8 tune 1.

  \item[Modelling of $\Vz$ reconstruction efficiency]
    The systematic uncertainty on the $\Vz$
    reconstruction efficiency is estimated from closure tests and from
    the stability of the results with respect to the $\Vz$
    selection cuts, as described in Section~\ref{ssec:strategy}.

  \item[Detector material] The overall mass of the tracker
    and the relative fractions of the different tracker materials
are varied in the simulations, with the requirement that the resulting predicted
tracker weight be consistent with the measured weight~\cite{CMS-NOTE-10-010}.
The difference between the results thus obtained and the nominal
results is taken as a contribution to the systematic uncertainty.

  \item[\GEANTfour cross sections] A 5\% systematic
    uncertainty is assigned to the baryon yields, as a result of the known
    imperfect modelling of the low-energy antiproton interaction cross
    section in the \GEANTfour version used~\cite{Aamodt:2010dx}.

  \item[Statistical uncertainty on the $\Vz$ yield correction] A small contribution to the total uncertainty stems from the finite size of the
    sample of minimum-bias events passed
    through the full detector simulation (50 million events), from which the correction is
    computed.
  \item[Estimation of $\Vz$ background]
    The uncertainty on the background remaining
    after $\Vz$ identification by means of the kinematic fit
    is taken
    as half the difference between the results of the two background
    estimation methods used.

\end{description}

The uncertainty on the beam spot position and size gives
a negligible contribution to the total uncertainty.

\begin{table}[htbp]
\begin{center}
\topcaption{Systematic uncertainties on the measured average $\Vz$ rates and $\PT$ sums.}
\begin{scotch}{l|r|r}
\multicolumn{3}{c}{\textbf{Average rates}} \\
\hline
 \textbf{Source} & $\PKzS$ (\%) & $\PgL$ (\%) \\
\hline
Leading charged-particle jet selection & 3 & 7 \\
Primary vertex selection & 1 & 1 \\
Modelling of $\Vz$ efficiency & & \\
\multicolumn{1}{r|}{charged-particle jet $\PT \leq 2.5\GeVc$} & 3 & 10 \\
\multicolumn{1}{r|}{charged-particle jet $\PT > 2.5\GeVc$} & 3 & 3 \\
Detector material & 3 & 3 \\
\GEANTfour cross sections & --- & 5 \\
Statistical uncertainty on $\Vz$ weights & & \\
\multicolumn{1}{r|}{$600\MeVc < \pt^{\Vz} < 700\MeVc$} & 0.1 & --- \\
\multicolumn{1}{r|}{$1.5\GeVc < \pt^{\Vz} < 1.6\GeVc$} & 0.03 & 0.33 \\
\multicolumn{1}{r|}{$6\GeVc < \pt^{\Vz}  < 8\GeVc$} & 1.4 & 8.3 \\
 Background estimation &  &  \\
\multicolumn{1}{r|}{charged-particle jet $\PT = 1\GeVc$} & 1.1 & 2 \\
\multicolumn{1}{r|}{charged-particle jet $\PT = 10\GeVc$} & 1.7 & 2 \\
\hline
\textbf{Total} &  &  \\
\multicolumn{1}{r|}{charged-particle jet $\PT = 1\GeVc$} & 6 & 14 \\
\multicolumn{1}{r|}{charged-particle jet $\PT = 10\GeVc$} & 6 & 10 \\
\hline
\multicolumn{3}{c}{\textbf{Average $\PT$ sums}} \\
\hline
 \textbf{Source} & $\PKzS$ (\%) & $\PgL$ (\%) \\
\hline
Background estimation &  &  \\
\multicolumn{1}{r|}{$\pt^{\Vz} = 600\MeVc$} & 0.1 & --- \\
\multicolumn{1}{r|}{$\pt^{\Vz} = 1.5\GeVc$} & 0.8 & 0.3 \\
\multicolumn{1}{r|}{$\pt^{\Vz}  = 8\GeVc$} & 3.6 & 4.0 \\
Other sources & as rates & as rates \\
\end{scotch}
\label{tab:syst}
\end{center}
\end{table}

\section{Results}
\label{sec:results}

\begin{figure}[htbp]
\begin{center}
\includegraphics[width=0.42\textwidth]{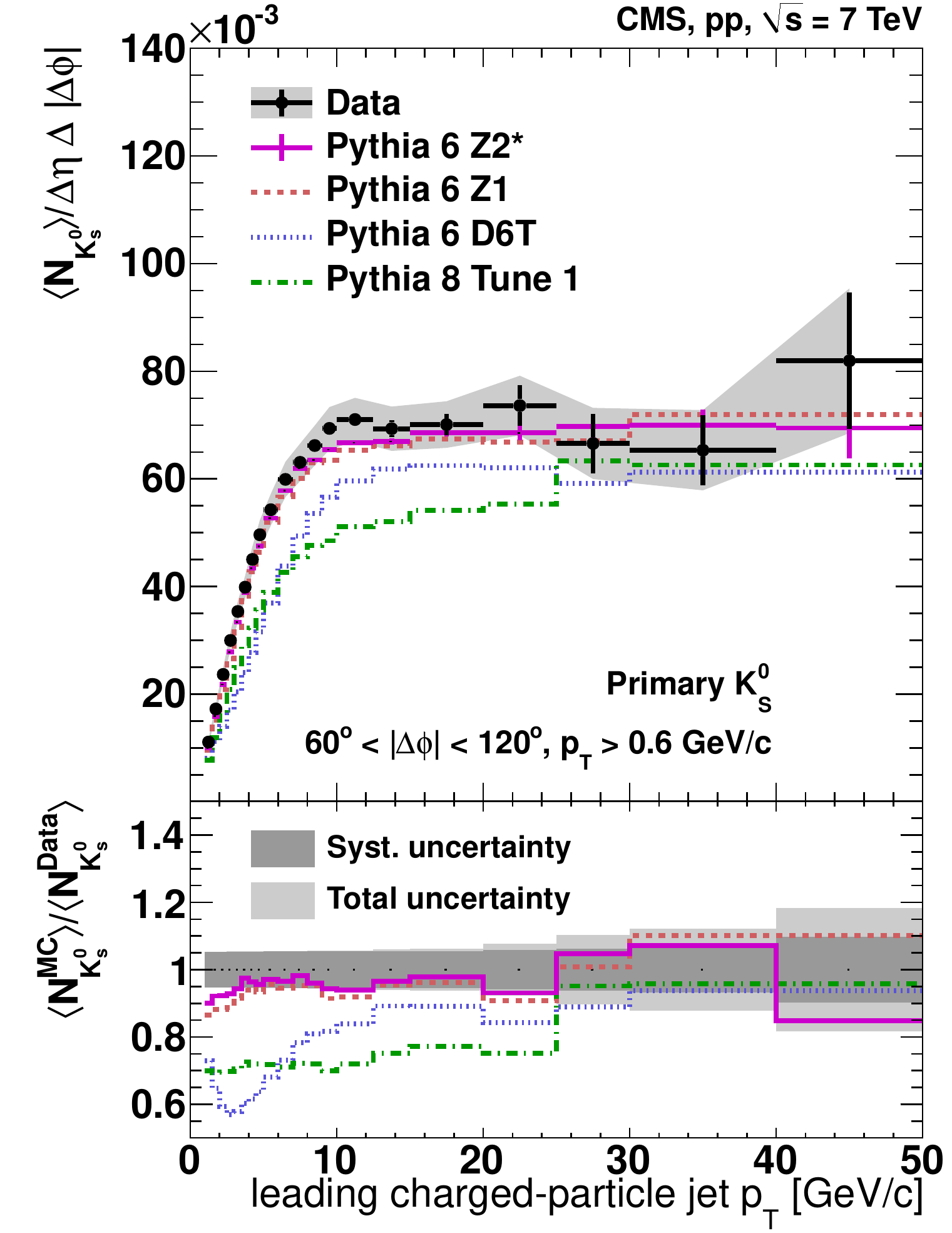}
\includegraphics[width=0.42\textwidth]{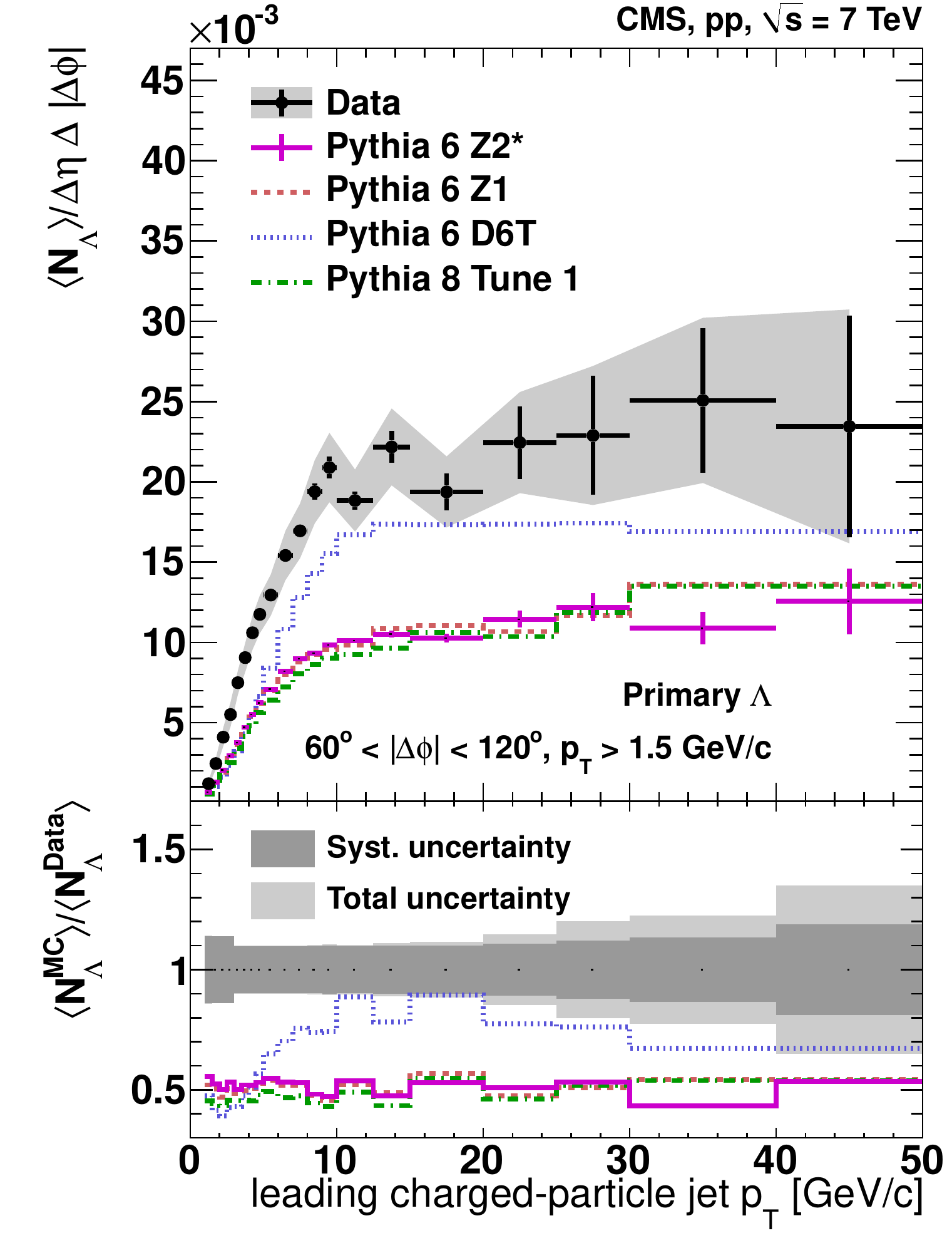}
\caption{Average multiplicity per unit of
  pseudorapidity and per radian in the transverse region ($\abs{\eta} < 2$, $60^\circ < \abs{\Delta \phi} < 120^\circ$), as a function
  of the $\PT$ of the leading charged-particle jet:
(\cmsLeft) $\PKzS$ with $\PT > 0.6\GeVc$; (\cmsRight) $\PgL$ with $\PT > 1.5\GeVc$.
Predictions of \PYTHIA tunes are compared to the data and the
ratios of simulations to data are shown in the bottom panels.
For the data, the statistical uncertainties (error bars) and the quadratic sum of statistical and systematic uncertainties (error band) are
shown, while for simulations the uncertainty is only
shown for \PYTHIA6 tune {Z2*}, for clarity.}
\label{fig:results_rates_pt}
\end{center}
\end{figure}

\begin{figure}[htbp]
\begin{center}
\includegraphics[width=0.42\textwidth]{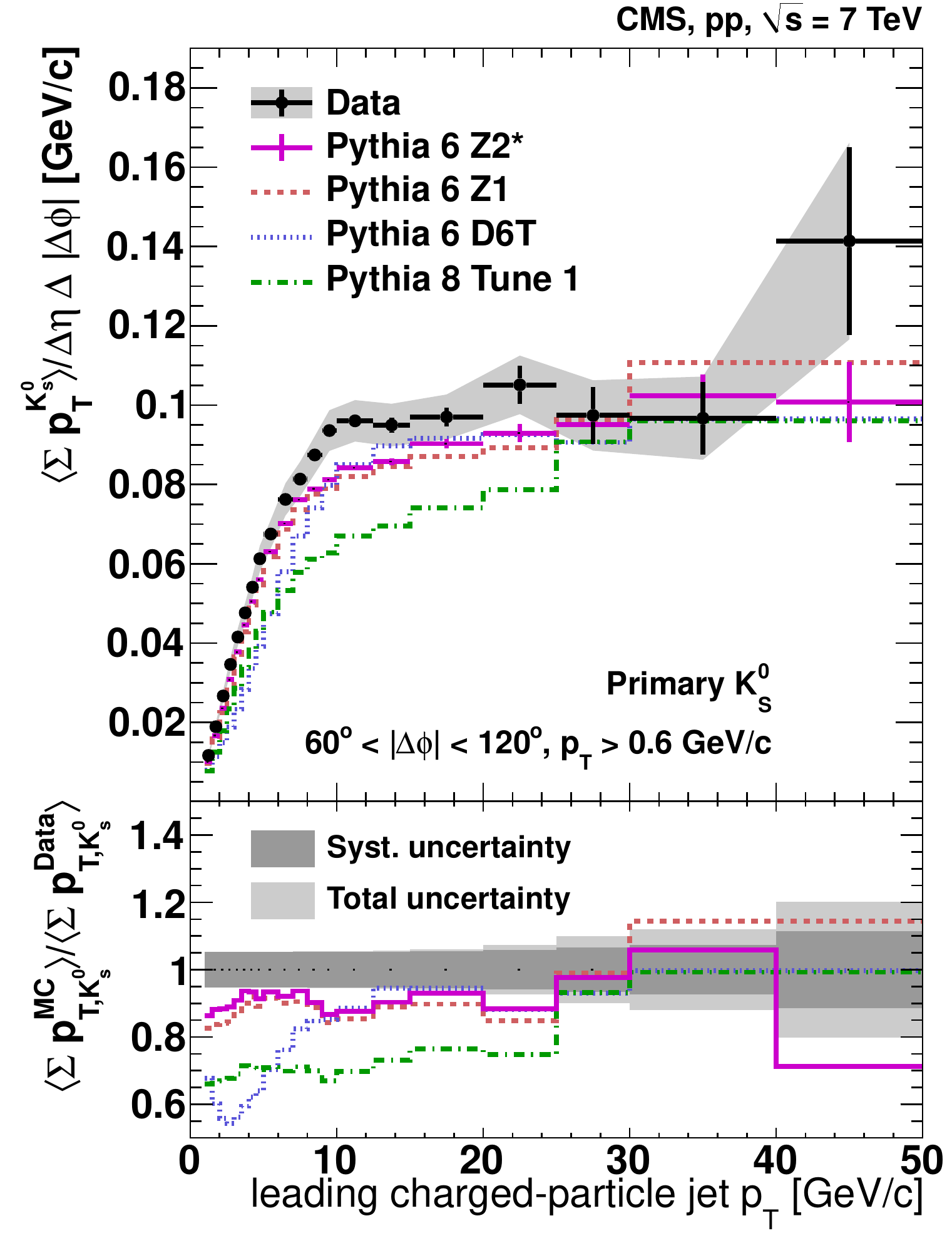}
\includegraphics[width=0.42\textwidth]{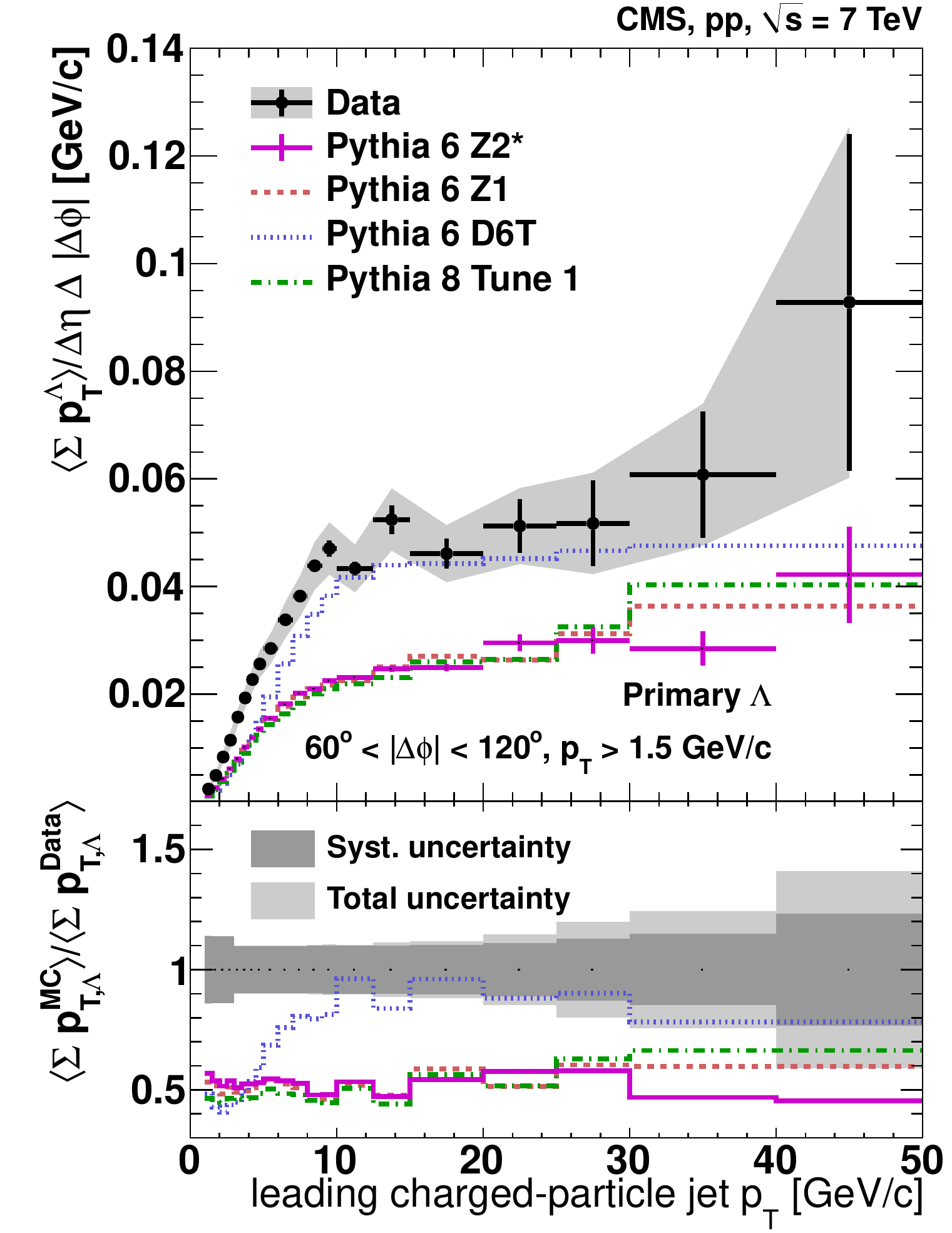}
\caption{Average scalar $\PT$ sum per unit of
  pseudorapidity and per radian in the transverse region ($\abs{\eta} < 2$, $60^\circ < \abs{\Delta \phi} < 120^\circ$), as a function
  of the $\PT$ of the leading charged-particle jet:
(\cmsLeft) $\PKzS$ with $\PT > 0.6\GeVc$; (\cmsRight) $\PgL$ with $\PT > 1.5\GeVc$.
Predictions of \PYTHIA tunes are compared to the data and the
ratios of simulations to data are shown in the bottom panels.
For the data, the statistical uncertainties (error bars) and the quadratic
sum of statistical and systematic uncertainties (error band) are
shown, while for simulations the uncertainty is only
shown for \PYTHIA6 tune {Z2*}, for clarity.}
\label{fig:results_sumpt}
\end{center}
\end{figure}

The $\Vz$ production rates in the transverse
region are shown in Fig.~\ref{fig:results_rates_pt}
as a function of the leading charged-particle jet $\PT$, and the $\Vz$ scalar $\PT$
sums in the transverse region are shown in
Fig.~\ref{fig:results_sumpt}.

The rates and $\PT$ sums exhibit a rise
with increasing hard scale, followed by a plateau.
The turn-on of the plateau is located at charged-particle jet $\PT \simeq 10\GeVc$
for both primary mesons and baryons.
Above the turn-on, the rates and $\PT$ sums are essentially constant,
implying also a constant strange-particle average $\PT$ above the turn-on.

A comparison can be made with the trends observed for charged primary
particles~\cite{CMS-PAS-QCD-10-010}
in spite of the different jet reconstruction
algorithm used in Ref.~\cite{CMS-PAS-QCD-10-010} (SISCone).
The dependence of the UE activity on the charged-particle jet $\PT$ is very similar to that observed for charged primary
particles~\cite{Khachatryan:2010pv, CMS-PAS-QCD-10-010, Aad:2010fh}.
The most striking feature is that the $\PT$ scale at which the
plateau starts, around $10\GeVc$ in $\Pp\Pp$ collisions at $\sqrt{s} = 7\TeV$, is independent of the type of primary
particle used to probe the UE activity.
These observations are consistent with the impact parameter picture of
particle production in hadron collisions~\cite{Sjostrand:1986ep,
  Frankfurt:2010ea},
in which the MPI contribution saturates at scales typical of central collisions.

The \PYTHIA6 Z1 and {Z2*} tunes qualitatively reproduce the
dependence of the $\PKzS$ rate
and $\PT$ sum on the charged-particle jet $\PT$, but exhibit a 10--15\% deficit in
the yield,
independent of the charged-particle jet $\PT$. \PYTHIA8 tune 1 underestimates the activity by about 30\%.
For the $\PgL$ baryons, \PYTHIA6 tunes Z1, {Z2*} and \PYTHIA8 tune 1
underestimate the rates by about 50\%.
After being tuned to the charged-particle data, \PYTHIA6 Z2* models
strangeness production in the UE in a very similar way as Z1,
in spite of the different parton distribution set used.

\begin{figure}[htbp]
\begin{center}
\includegraphics[width=0.45\textwidth]{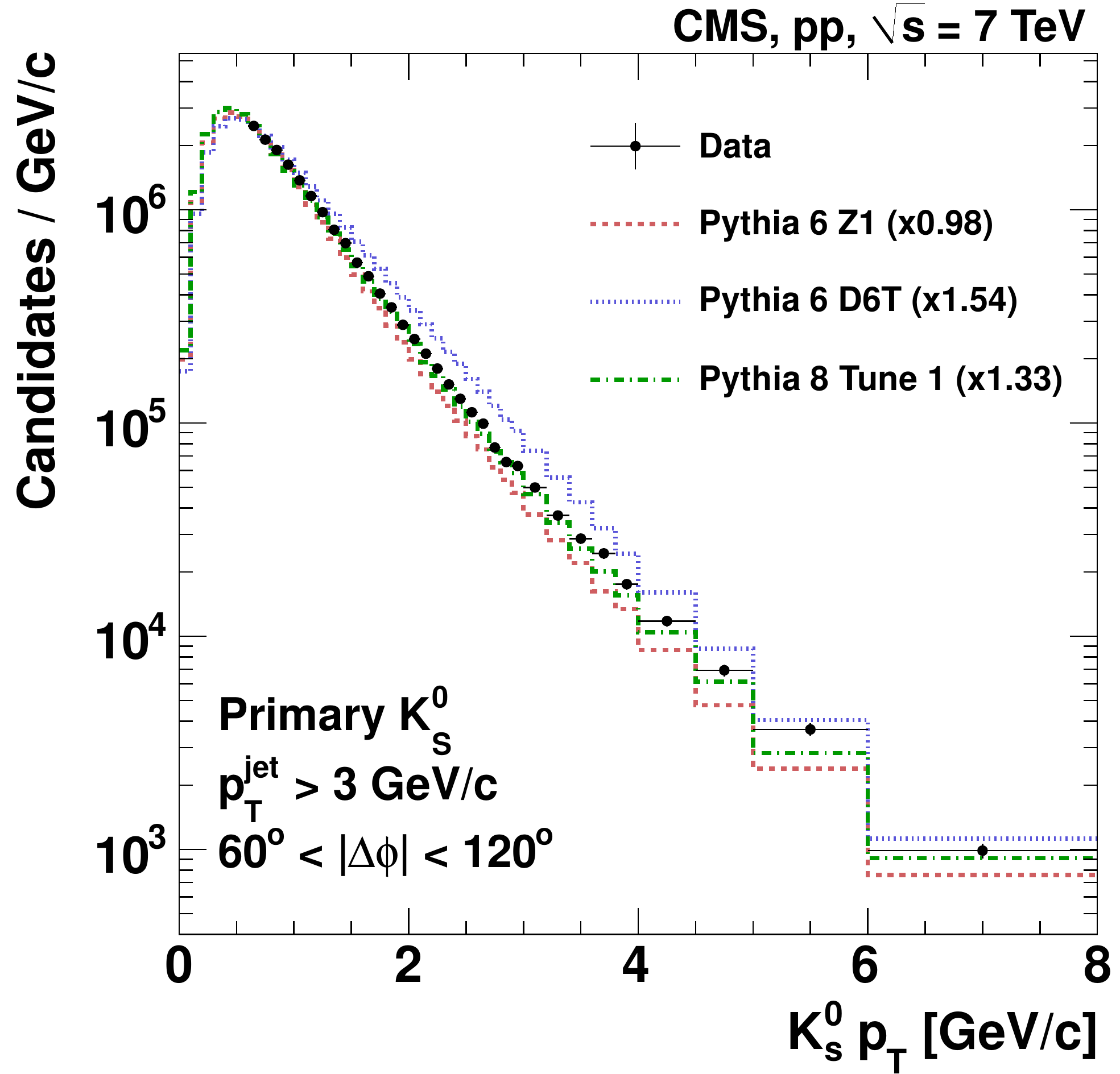}
\includegraphics[width=0.45\textwidth]{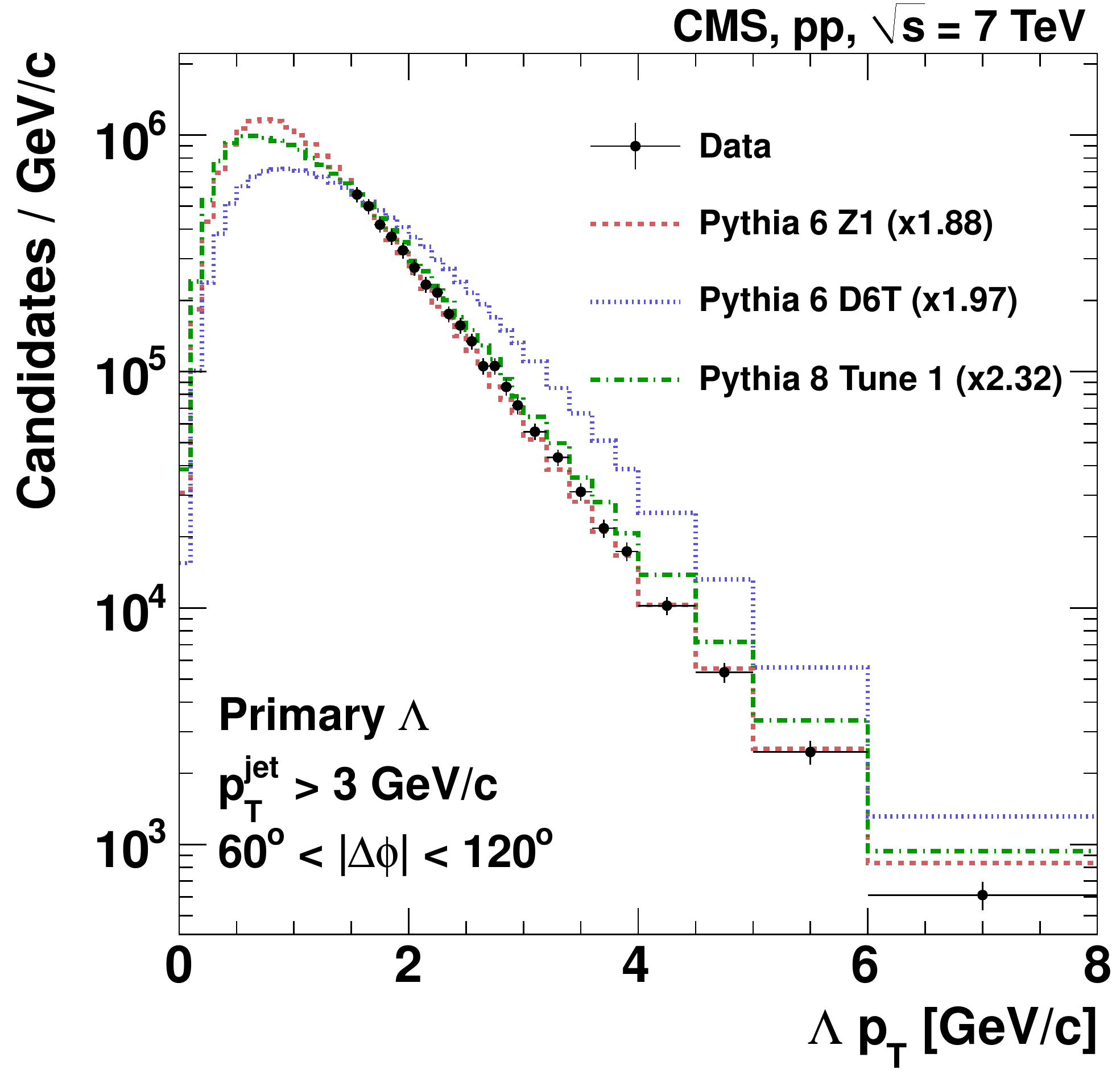}
\caption{$\Vz$ $\PT$ distributions corrected for
  selection efficiency and background without
  correction to the leading charged hadron jet,
  in the region transverse
  to a leading reconstructed charged-particle jet with $\PT > 3\GeVc$,
  compared to predictions from different \PYTHIA tunes (\cmsLeft: $\PKzS$;
  \cmsRight: $\PgL$). Error bars indicate the quadratic sum
  of the statistical and systematic uncertainties. Simulations are normalized to the
  first $\PT$ bin in the data, with normalization factors given in
  parentheses.}
\label{fig:v0corr_pt}
\end{center}
\end{figure}

\PYTHIA6 D6T shows a dependence of the activity
on the charged-particle jet $\PT$ that differs from that of the data
and of the other tunes.
In addition, the $\Vz$ $\PT$ distributions predicted by
\PYTHIA6 D6T in the transverse region are
in strong disagreement with the data. As an illustration,
the $\PT$ spectra are shown in Fig.~\ref{fig:v0corr_pt} for events
with a reconstructed charged-particle jet $\PT > 3\GeVc$
(without correction to the leading charged hadron jet).
For the $\PKzS$ case, in the $\PT$ range observed ($\PT > 600\MeVc$),
\PYTHIA6 tune {D6T} shows a
much harder spectrum than the data, while tune {Z1} shows a
softer spectrum and \PYTHIA8 tune 1 reproduces
the shape well. For the $\PgL$ case,
in the $\PT > 1.5\GeVc$ range, \PYTHIA6 D6T shows a much harder spectrum than
the data, while the other simulations describe the data reasonably
well.

\begin{figure*}[htbp]
\begin{center}
\includegraphics[width=0.45\textwidth]{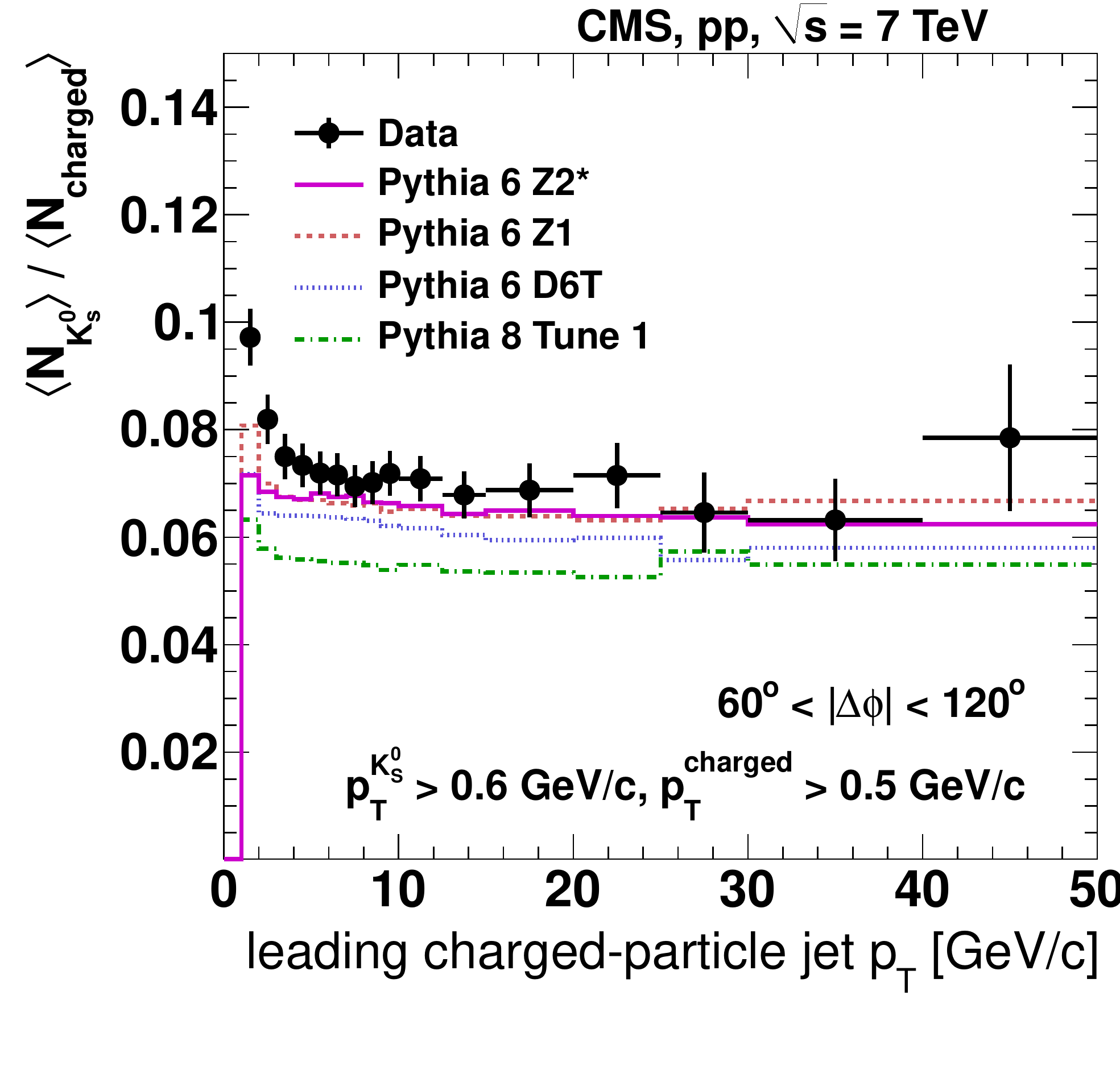}
\includegraphics[width=0.45\textwidth]{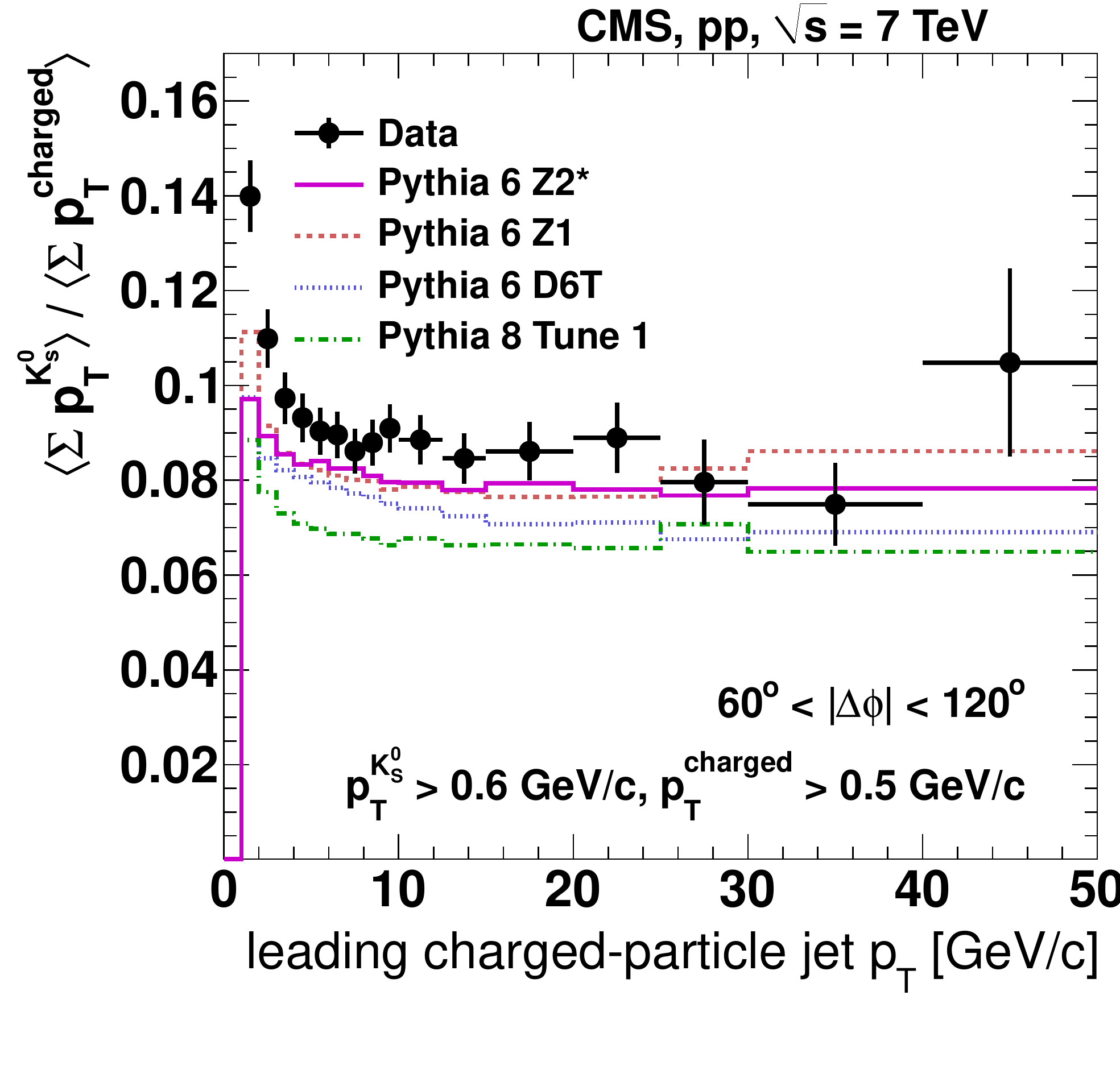}
\includegraphics[width=0.45\textwidth]{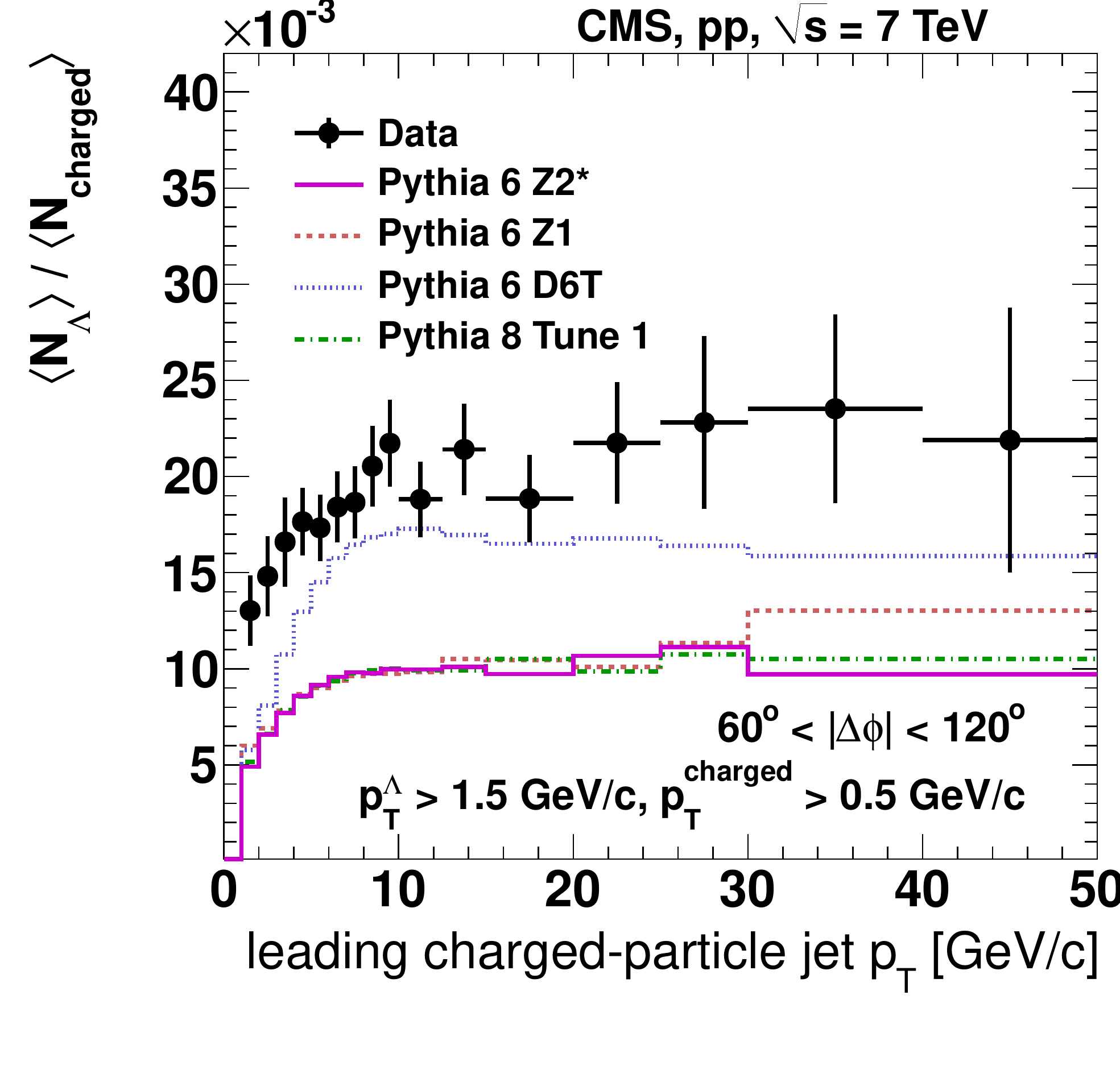}
\includegraphics[width=0.45\textwidth]{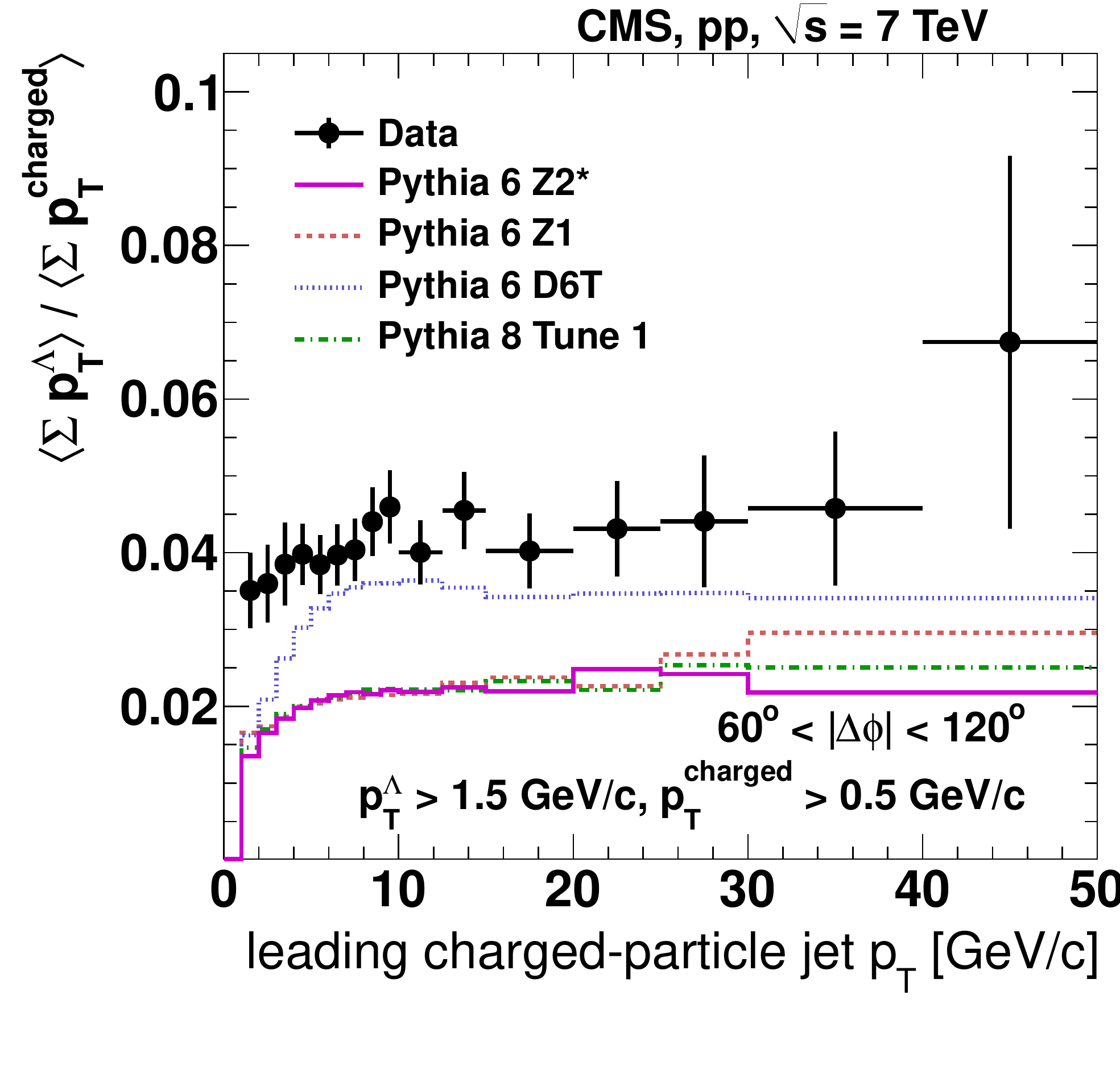}
\caption{Ratios of the average multiplicities and scalar $\PT$ sums for primary
$\Vz$ in the transverse region to the same quantities for
primary charged particles~\cite{CMS-PAS-QCD-10-010}
as a function of charged-particle jet $\PT$.
The error bars indicate the quadratic sum of
the statistical and systematic uncertainties.}
\label{fig:ratios}
\end{center}
\end{figure*}

The ratios of the rates and $\PT$ sums of primary
$\Vz$ mesons to the rates and $\PT$ sums of primary charged
particles from Ref.~\cite{CMS-PAS-QCD-10-010} are shown in
Fig.~\ref{fig:ratios}. The data are integrated over the same
pseudorapidity range for strange and charged particles, $\abs{\eta} < 2$.
The $\PKzS$ to charged-particle activity ratios are constant in the charged-particle jet $\PT$ range 3--50\GeVc, \ie almost
throughout the whole range studied and, specifically, across the turn-on of
the plateau around $10\GeVc$.
An increase is seen below $3\GeVc$.
This feature is also present in the simulations but is not as pronounced
as in the data, and not in all tunes studied.

The $\PgL$ to charged-particle activity ratios exhibit a
rise for charged-particle jet $\PT< 10\GeVc$, followed by a plateau. A similar
dependence is visible in \PYTHIA. Simulations indicate that the rise is related to the
observed hardening of the baryon $\PT$ spectrum
as the charged-particle jet $\PT$ increases, combined with the $1.5\GeVc$ $\PT$ cut
applied to the baryon sample. When the baryon $\PT$ cut is decreased to $0.5\GeVc$ as for charged particles, constant ratios are predicted.

Constant strange- to charged-particle activity ratios have thus been
measured for $\PKzS$ mesons for charged-particle jet $\PT > 3\GeVc$ and for $\PgL$ baryons
for charged-particle jet $\PT > 10\GeVc$.
In addition, as just discussed, when accounting for the acceptance of the
baryon $\PT$ cut, a constant ratio is also predicted for $\PgL$ baryons at
charged-particle jet $\PT < 10\GeVc$.
Since the trends observed are very similar for charged and
strange particles, as well as for mesons and baryons, the present
measurements suggest that hadronization and MPI are decoupled.

\section{Conclusions}
\label{sec:conclusions}

This paper describes measurements of the underlying event activity
in $\Pp\Pp$ collisions at $\sqrt{s} = 7\TeV$,
probed through the production of primary $\PKzS$ mesons and $\PgL$ baryons.
The production of $\PKzS$ mesons and $\PgL$
baryons in the kinematic range $\pt^{\PKzS} > 0.6\GeVc$,
$\pt^{\PgL} > 1.5\GeVc$ and $\abs{\eta} < 2$
is analysed in the transverse region, defined as $60^\circ<
\abs{\Delta\phi} < 120^\circ$, with $\Delta\phi$ the difference
in azimuthal angle between the leading charged-particle jet and the strange particle
directions.
The average multiplicity and the average scalar $\PT$ sum of primary particles per event
are studied as a function of the leading charged-particle jet $\PT$.

A steep rise of the underlying event activity is seen
with increasing leading jet $\PT$, followed
by a ``saturation'' region for jet $\PT > 10\GeVc$.
This trend and the $\PT$ scale above which saturation occurs
are very similar to those observed with
charged primary particles. The similarity of the behaviour for strange and
charged particles is consistent with the impact-parameter
picture of multiple parton interactions in $\Pp\Pp$ collisions, in which the
centrality of the $\Pp\Pp$ collision
and the MPI activity are correlated.

The results are compared to recent tunes of the \PYTHIA Monte Carlo event generator. The \PYTHIA
simulations underestimate the data by 15--30\% for
$\PKzS$ mesons and by about 50\% for $\PgL$
baryons, a MC deficit similar to that observed for the inclusive strange particle production in $\Pp\Pp$ collisions.

The constant strange- to charged-particle activity ratios and the similar trends for
mesons and baryons indicate that the MPI dynamics is decoupled from parton
hadronization, with the latter occurring at a later stage.

\section*{Acknowledgments}

\hyphenation{Bundes-ministerium Forschungs-gemeinschaft Forschungs-zentren} We congratulate our colleagues in the CERN accelerator departments for the excellent performance of the LHC and thank the technical and administrative staffs at CERN and at other CMS institutes for their contributions to the success of the CMS effort. In addition, we gratefully acknowledge the computing centres and personnel of the Worldwide LHC Computing Grid for delivering so effectively the computing infrastructure essential to our analyses. Finally, we acknowledge the enduring support for the construction and operation of the LHC and the CMS detector provided by the following funding agencies: the Austrian Federal Ministry of Science and Research and the Austrian Science Fund; the Belgian Fonds de la Recherche Scientifique, and Fonds voor Wetenschappelijk Onderzoek; the Brazilian Funding Agencies (CNPq, CAPES, FAPERJ, and FAPESP); the Bulgarian Ministry of Education, Youth and Science; CERN; the Chinese Academy of Sciences, Ministry of Science and Technology, and National Natural Science Foundation of China; the Colombian Funding Agency (COLCIENCIAS); the Croatian Ministry of Science, Education and Sport; the Research Promotion Foundation, Cyprus; the Ministry of Education and Research, Recurrent financing contract SF0690030s09 and European Regional Development Fund, Estonia; the Academy of Finland, Finnish Ministry of Education and Culture, and Helsinki Institute of Physics; the Institut National de Physique Nucl\'eaire et de Physique des Particules~/~CNRS, and Commissariat \`a l'\'Energie Atomique et aux \'Energies Alternatives~/~CEA, France; the Bundesministerium f\"ur Bildung und Forschung, Deutsche Forschungsgemeinschaft, and Helmholtz-Gemeinschaft Deutscher Forschungszentren, Germany; the General Secretariat for Research and Technology, Greece; the National Scientific Research Foundation, and National Office for Research and Technology, Hungary; the Department of Atomic Energy and the Department of Science and Technology, India; the Institute for Studies in Theoretical Physics and Mathematics, Iran; the Science Foundation, Ireland; the Istituto Nazionale di Fisica Nucleare, Italy; the Korean Ministry of Education, Science and Technology and the World Class University program of NRF, Republic of Korea; the Lithuanian Academy of Sciences; the Mexican Funding Agencies (CINVESTAV, CONACYT, SEP, and UASLP-FAI); the Ministry of Science and Innovation, New Zealand; the Pakistan Atomic Energy Commission; the Ministry of Science and Higher Education and the National Science Centre, Poland; the Funda\c{c}\~ao para a Ci\^encia e a Tecnologia, Portugal; JINR (Armenia, Belarus, Georgia, Ukraine, Uzbekistan); the Ministry of Education and Science of the Russian Federation, the Federal Agency of Atomic Energy of the Russian Federation, Russian Academy of Sciences, and the Russian Foundation for Basic Research; the Ministry of Science and Technological Development of Serbia; the Secretar\'{\i}a de Estado de Investigaci\'on, Desarrollo e Innovaci\'on and Programa Consolider-Ingenio 2010, Spain; the Swiss Funding Agencies (ETH Board, ETH Zurich, PSI, SNF, UniZH, Canton Zurich, and SER); the National Science Council, Taipei; the Thailand Center of Excellence in Physics, the Institute for the Promotion of Teaching Science and Technology of Thailand and the National Science and Technology Development Agency of Thailand; the Scientific and Technical Research Council of Turkey, and Turkish Atomic Energy Authority; the Science and Technology Facilities Council, UK; the US Department of Energy, and the US National Science Foundation.

Individuals have received support from the Marie-Curie programme and the European Research Council and EPLANET (European Union); the Leventis Foundation; the A. P. Sloan Foundation; the Alexander von Humboldt Foundation; the Belgian Federal Science Policy Office; the Fonds pour la Formation \`a la Recherche dans l'Industrie et dans l'Agriculture (FRIA-Belgium); the Agentschap voor Innovatie door Wetenschap en Technologie (IWT-Belgium); the Ministry of Education, Youth and Sports (MEYS) of Czech Republic; the Council of Science and Industrial Research, India; the Compagnia di San Paolo (Torino); the HOMING PLUS programme of Foundation for Polish Science, cofinanced by EU, Regional Development Fund; and the Thalis and Aristeia programmes cofinanced by EU-ESF and the Greek NSRF.

\bibliography{auto_generated}

\cleardoublepage \appendix\section{The CMS Collaboration \label{app:collab}}\begin{sloppypar}\hyphenpenalty=5000\widowpenalty=500\clubpenalty=5000\textbf{Yerevan Physics Institute,  Yerevan,  Armenia}\\*[0pt]
S.~Chatrchyan, V.~Khachatryan, A.M.~Sirunyan, A.~Tumasyan
\vskip\cmsinstskip
\textbf{Institut f\"{u}r Hochenergiephysik der OeAW,  Wien,  Austria}\\*[0pt]
W.~Adam, T.~Bergauer, M.~Dragicevic, J.~Er\"{o}, C.~Fabjan\cmsAuthorMark{1}, M.~Friedl, R.~Fr\"{u}hwirth\cmsAuthorMark{1}, V.M.~Ghete, N.~H\"{o}rmann, J.~Hrubec, M.~Jeitler\cmsAuthorMark{1}, W.~Kiesenhofer, V.~Kn\"{u}nz, M.~Krammer\cmsAuthorMark{1}, I.~Kr\"{a}tschmer, D.~Liko, I.~Mikulec, D.~Rabady\cmsAuthorMark{2}, B.~Rahbaran, C.~Rohringer, H.~Rohringer, R.~Sch\"{o}fbeck, J.~Strauss, A.~Taurok, W.~Treberer-Treberspurg, W.~Waltenberger, C.-E.~Wulz\cmsAuthorMark{1}
% ...
\vskip\cmsinstskip
% ...
1:~~Also at Vienna University of Technology, Vienna, Austria\\
2:~~Also at CERN, European Organization for Nuclear Research, Geneva, Switzerland\\
% ...

\end{sloppypar}
\end{document}